\documentclass[a4paper,12pt]{article}
\pdfoutput=1
\usepackage{epsfig,graphicx,xcolor,soul,amsbsy,amssymb,latexsym,amsfonts,amsmath,tcolorbox,setspace}
\usepackage{pstricks}
\usepackage{color}
\usepackage[numbers,square,comma, compress]{natbib}

\usepackage{eurosym}
\usepackage[a4paper]{geometry}
\usepackage{graphicx}
\usepackage{subfigure}
\usepackage{tikz}
\usepackage{xcolor}
\usepackage{amsmath,amssymb,amsfonts,tcolorbox,pstricks,setspace}

\numberwithin{equation}{section}

\newcommand{\bea}{\begin{eqnarray}\displaystyle}
\newcommand{\eea}{\end{eqnarray}}
\newcommand{\nn}{\nonumber \\}

\newcommand{\figref}[1]{Fig.~\protect\ref{#1}}

\title{\begin{flushright}{\vspace{-2.5cm}\small SNUST 15-07}\end{flushright}\vspace{0.8cm}
\bf{ Instanton-Monopole Correspondence from\\
M-Branes on $\mathbb{S}^1$ and Little String Theory}\\[15pt]}
\author{\large \textsc{Stefan~Hohenegger\footnote{\tt s.hohenegger@ipnl.in2p3.fr}~,~Amer Iqbal\footnote{\tt  amer@alum.mit.edu}~,~ Soo-Jong Rey\,\footnote{\tt sjrey@snu.ac.kr}}}
\date{}

\begin{document}

\maketitle

\begin{center}
\renewcommand{\thefootnote}{\fnsymbol{footnote}}\vspace{-0.5cm}
${}^{\footnotemark[1]}$ Universit\'e Claude Bernard (Lyon 1)\\UMR 5822, CNRS/IN2P3, Institut de Physique Nucl\'eaire, Bat. P. Dirac\\ 4 rue Enrico Fermi, F-69622-Villeurbanne, \rm FRANCE\\[0.4cm]
${}^{\footnotemark[2]}$ Abdus Salam School of Mathematical Sciences \\ Government College University, Lahore, Pakistan\\[0.4cm]
${}^{\footnotemark[2]}$ Center of Mathematical Sciences and Applications, Harvard University\\ 1 Oxford Street, Cambridge, MA 02138, USA\\[0.4cm]
${}^{\footnotemark[3]}$ School of Physics and Astronomy \& Center for Theoretical Physics\\
          Seoul National University, Seoul 08826 \rm KOREA\\[0.4cm]
${}^{\footnotemark[3]}$ Fields, Gravity \& Strings, Center for Theoretical Physics of the Universe \\
Institute for Basic Sciences, Daejeon 34047 \rm KOREA\\[1cm]
\end{center}

\begin{abstract}
\noindent
We study BPS excitations in M5-M2-brane configurations with a compact transverse direction, which are also relevant for type IIa and IIb little string theories. These configurations are dual to a class of toric elliptically fibered Calabi-Yau manifolds $X_N$ with manifest $SL(2,\mathbb{Z})\times SL(2,\mathbb{Z})$ modular symmetry. They admit two dual gauge theory descriptions. For both, the  non-perturbative partition function can be written as an expansion of the topological string partition function of $X_N$ with respect to either of the two modular parameters. We analyze the resulting BPS counting functions in detail and find that they can be fully constructed as linear combinations of the  BPS counting functions of M5-M2-brane configurations with non-compact transverse directions. For certain M2-brane configurations, we also find that the free energies in the two dual theories agree with each other, which points to a new correspondence between instanton and monopole configurations. These results are also a manifestation of T-duality between type IIa and IIb little string theories.

\end{abstract}

${}$\\[500pt]

\tableofcontents

\onehalfspacing
\newpage
\vskip1cm

\rightline{\sl If you do not expect the unexpected, you will not find it;}
\rightline{\sl for it is hard to be sought out and difficult.}
\rightline{---\textsc{heraclitus}}
\section{Introduction and Summary}
In recent years, the interplay between M-theory/string-theory, geometry and superconformal gauge theories has been rigorously studied, leading to new and deep insights. At the focus of interest are configurations of $N$ parallel M5-branes with multiple M2-branes stretched between them (see \emph{e.g.} \cite{Haghighat:2013gba,Haghighat:2013tka,
Hohenegger:2013ala,
Haghighat:2015coa,Hohenegger:2015cba,
Haghighat:2015ega}). These brane configurations are known to be U-dual to specific toric elliptically fibered Calabi-Yau threefolds $X_N$ over an $A_{N-1}$ base space. They can also be associated to six-dimensional nonabelian supersymmetric field theories, which upon further compactification to four-dimensions give rise to mass-deformed $\mathcal{N}=2^*$ gauge theories. All these six-dimensional systems exhibit very rich dynamics and contain extended BPS degrees of freedom that are unfamiliar from a four-dimensional point of view.

Indeed, as was first pointed out in \cite{Haghighat:2013gba}, the configuration of M2-branes stretched between M5-branes described above gives rise to one-dimensional dynamical objects at the brane intersections. When the M5-branes coincide, these so-called {\sl M-strings} become tensionless, forming essential interacting degrees of freedom of the elusive (2,0) superconformal, local quantum field theory. When the M5-branes are separated, the M-strings become BPS string states with tension. Their BPS excitations, which are expected to elucidate the worldsheet dynamics over the six-dimensional target space, are counted by the topological string partition function of the dual toric Calabi-Yau manifold $X_N$ \cite{Haghighat:2013gba,Haghighat:2013tka,Hohenegger:2013ala}. This partition function is efficiently computed by the refined topological vertex approach \cite{TV,Hollowood:2003cv,Iqbal:2007ii}, and depends on two parameters, $\epsilon_{1,2}$ which are fugacities for the little group $SO(4)$ of massive particles in five dimensions in M-theory compactification on a Calabi-Yau threefold. From the viewpoint of the non-perturbative gauge theory partition function, these parameters correspond to putting the gauge theory on a curved spacetime, the so-called generalized $\Omega$-background \cite{Nekrasov:2002qd}. 

Another manifestation of string degrees of freedom was discussed in \cite{Haghighat:2015coa,Hohenegger:2015cba}: upon compactification to five dimensions, the M-strings become (electrically charged) BPS particles which are related via five-dimensional S-duality to magnetically charged monopole strings. While the precise details of this duality map are somewhat intricate \cite{Douglas:2010iu}, we proposed in \cite{Hohenegger:2015cba} that the degeneracies of certain M-string BPS-configurations captures the elliptic genus (see \cite{EG} for its general definition) of the moduli space of monopole strings. This proposal applies to theories of $SU(N)$ gauge theories for any $N$ and for general distributions of the constituent monopole strings. In \cite{Hohenegger:2015cba}, we successfully checked this proposal for all known cases, namely, the Taub-NUT and Atiyah-Hitchin spaces. These spaces correspond to the moduli spaces of charge $(1,1)$ monopoles in $SU(3)$ \cite{taub-nut} and charge $(2)$ monopoles in $SU(2)$ gauge theory \cite{AtiyahHitchin}, respectively. The elliptic genera of their respective moduli spaces were previously computed in \cite{Harvey:2014nha,Bak:2014xwa}. In \cite{Hohenegger:2015cba}, we studied the elliptic genus of the moduli space of monopole strings for arbitrary gauge group and for general distribution of constituent monopole strings.

The purpose of this paper is to expose new phenomena associated with a richer duality structure that arises when the above M5-M2 brane setup is extended to a configuration with a larger modular symmetry group. Such an extension appears in a variety of physical problems. We focus on a particularly interesting configuration that has to do with compactifying a direction transverse to the M5-branes to a circle. Concretely, the brane configuration studied in \cite{Hohenegger:2015cba} consists of $N$ parallel M5-branes which are separated along a non-compact direction. Here, we compactify this direction to $\mathbb{S}^1$. Geometrically, this modified brane configuration is again dual to a toric elliptically fibered Calabi-Yau threefold $X_N$. However, in contrast to the non-compact case, the base is now an {\sl affine} $A_{N-1}$ space, which in turn is a fibration over $\mathbb{P}^1$. As a consequence of this two-fold fibration structure, this setup exhibits manifest $SL(2,\mathbb{Z})\times SL(2,\mathbb{Z})$ symmetry. This two-fold $SL(2, \mathbb{Z})$ symmetry permits to describe this theory by using two different approaches:
\begin{itemize}
\item The first approach relates the compact brane configuration to two different gauge theories: theory 1 is the Coulomb branch of a $U(N)$ gauge theory, while theory 2 is a circular quiver with $N$ nodes of $U(1)$ gauge theories. At a generic value of the parameters, both are $[U(1)]^N$ circular quiver gauge theories. The difference is that, when the M5-branes are all separated, theory 1 has massive bifundamentals, while theory 2 has massless bifundamentals. The two gauge theories arise from the map of the M-theory brane configuration to Type IIB brane configurations consisting of either one NS5-brane and $N$ D5-branes or one D5-brane and $N$ NS5-branes, intersecting in both cases on a torus. The former gives rise to the theory 1, while the latter gives rise to the theory 2.  Therefore, the two gauge theories are related to each other by Type IIB S-duality. On the other hand, in the description in terms of the toric Calabi-Yau manifold $X_N$, the two gauge theories are just two facets of topological string theory and are related to each other by an exchange of the base and the fiber in $X_N$. 
As such, the partition functions of these two gauge theories can be extracted from the topological string partition function of $X_N$ by expanding in two different parameters. These correspond to the two modular parameters of $SL(2,\mathbb{Z})\times SL(2,\mathbb{Z})$ mentioned above.

\item The second approach relates the compact brane configuration to maximally supersymmetric little string theories in six dimensions \cite{Berkooz:1997cq,Seiberg:1997zk, Losev:1997hx, 
Aharony:1998ub} \footnote{See \cite{Aharony:1999ks}, \cite{Kutasov:2001uf} for reviews of little string theories on $\mathbb{R}^{5,1}$ and \cite{Aharony:2015zea} for little string theories on $AdS_5 \times \mathbb{S}^1$. }. These little strings are fundamental strings bound to NS5-branes that are decoupled from the ambient ten-dimensional spacetime. Therefore, descending from NS5-branes in Type IIA and IIB string theories or ALE singularities in Type IIB and IIA string theories, there are type IIb and IIa little string theories in six dimensions with $(2,0)$ and $(1,1)$ supersymmetries, respectively \footnote{Our notations adhere to the convention that non-chiral string theories are labelled as A or a, while chiral string theories are labelled as B or b. We trust this will cause maximal confusion to the readers.}. These little string theories are nonlocal theories since excitations contain ``little strings" of finite tension. In the brane configuration description, the $\mathbb{S}^1$ compactification transverse to M5-branes renders the tension of these little strings. Moreover, one can see from U-duality of the brane configuration that the gauge theory 1 and gauge theory 2 are related to Type IIa and IIb little string theory, respectively. In the same way as the two gauge theories are related to each other by the exchange of the two coupling parameters, upon compactification on $\mathbb{S}^1$, the IIa and IIb little string theories are T-dual to each other by the exchange of their $SL(2, \mathbb{Z}) \times SL(2, \mathbb{Z})$ modular parameters.
\footnote{We thank the authors of \cite{Bhardwaj:2015oru}, communicated through Cumrun Vafa, for suggesting possible relations between these two approaches.}
\end{itemize}
We analyze the modular properties of the partition functions of the two pairs of dual theories mentioned above and discover two remarkable properties. First of all, the functions capturing the degeneracies of single particle BPS states of gauge theory 2 can be expressed by the analog functions of degeneracies of monopole strings in the non-compact M5-brane configuration as worked out in \cite{Hohenegger:2015cba}. Roughly speaking, the free energy of compact monopole strings can be expressed as a linear combination of the free energies of non-compact monopole strings. Secondly, the generating functions of degeneracies for certain instanton configurations of theory 1 are equal to the generating functions of degeneracies for monopole strings of theory 2. The equivalence we observe is case-specific in the sense that it maps configurations which are fully covariant under the respective $SL(2,\mathbb{Z})$ symmetries into each other. A more careful study of the relation of the remaining configurations (and thus a possible equivalence of the two partition functions) is currently under way~\cite{ours}.

This paper is organized as follows. In section~\ref{Sect:BraneConfiguration}, we discuss in detail the M-brane configuration and the dual Calabi-Yau threefold $X_N$. We also describe the two distinct gauge theories associated to $X_N$, and relate them to type IIa and IIb little string theories. In section~\ref{Sect:PartitionFunction}, we present the topological partition function $\mathcal{Z}_{X_N}$ of $X_N$ and discuss in detail the manifest $SL(2,\mathbb{Z})\times SL(2,\mathbb{Z})$ modular symmetry, in particular the transformation properties of $\mathcal{Z}_{X_N}$. Furthermore, we extract the non-perturbative partition functions of the two gauge theories mentioned above by expanding them in the parameters associated with the two different $SL(2,\mathbb{Z})$'s. In section~\ref{Sect:RelCompNonComp}, we find that the gauge theory free-energies can be expressed in terms of their non-compact counterparts that we analyzed in the previous work~\cite{Hohenegger:2015cba}.
In section~\ref{Sect:MonopolesvsInst}, we exhibit remarkable relations between the free energies of the two different gauge theories. These relations are very non-trivial in that they relate quantities computed in the instanton moduli space with counting functions of multi-monopole string configurations. In section~\ref{Sect:EllGen}, following the conjecture in \cite{Hohenegger:2015cba} for the non-compact case, we propose a concrete expression for the elliptic genus of the monopole moduli space of the affine ${A}_{N-1}$ theory. From this, we extract the corresponding $\chi_y$-genus which encodes topological invariants of this moduli space. We conclude in section~\ref{Sect:Conclusions} and point out further directions for future research. Appendix~\ref{App:PertExpand} contains explicit series expansions of BPS-counting functions of various instanton and monopole configurations.

\section{Brane Configuration on $\mathbb{S}^1$ and Dual Theories}
\label{Sect:BraneConfiguration}

\subsection{M-Brane Configuration}\label{Sect:BraneConfig}
Our starting point is a particular BPS configuration of M-branes in the eleven-dimensional M-theory vacuum $\mathbb{T}^2\times \mathbb{R}_{\parallel}^3\times \mathbb{S}^1\times \mathbb{S}^1\times \mathbb{R}_\perp^4$ (with $\mathbb{T}^2\sim\mathbb{S}^1\times \mathbb{S}^1$), parametrized by the Cartesian coordinates $(x^0,\ldots,x^{10})$. Specifically, we consider $N$ planar M5-branes, $K$ open M2-branes stretched between M5-branes, and $M$ M-waves on two-dimensional intersection of M5-branes and M2-branes. The precise configuration is summarized in the following table
\begin{align}
\label{braneconfig}
&\begin{array}{|c|cc|ccc|c|c|cccc|}\hline
&  x^0 & x^1& x^2 & x^3 & x^4 & \,x^5\, &\, x^6 &x^7&x^8&x^9&x^{10} \\
\hline
M5 \ & = & =  & = &=&=&=&&&&& \\
M2 \ & = & =& & & & & = &&&& \\
M\sim & = & =& && & &&&&& \\\hline
\end{array}\\[-14pt]
&\hspace{1.2cm}\underbrace{\hspace{1.4cm}}_{\mathbb T^2\sim \mathbb{S}^1\times \mathbb{S}^1}\,\underbrace{\hspace{2.1cm}}_{\mathbb{R}_{\parallel}^3}\,\underbrace{\hspace{0.1cm}}_{\mathbb{S}^1_{R_5}}\,\underbrace{\hspace{0.1cm}}_{\mathbb{S}^1_{R_6}}\,\underbrace{\hspace{3cm}}_{\mathbb{R}_\perp^4}\nonumber
\end{align}
This brane configuration is very similar to the one studied in \cite{Hohenegger:2015cba}, 
with the only difference that, in the present setup, in addition to $x^1 \simeq x^1 + 2 \pi R_1$, the direction $x^6\simeq x^6+2\pi R_6$ is compactified to a circle with radius $R_6$. 
{ The open M2-branes are extended along $\mathbb{S}^1_{R_0} \times \mathbb{S}^1_{R_1} \times \mathbb{S}^1_{R_6}$. We denote the geometric parameters of this $\mathbb{T}^3$ as 
\begin{align}
&2 \pi i R_1 := \tau && \mbox{and} && 
2 \pi i R_6 := \rho
\label{tau-rho}
\end{align}
and their respective fugacities as 
\begin{align}
&Q_\tau = e^{2 \pi i \tau} && \mbox{and} && Q_\rho = e^{2 \pi i \rho}.
\nonumber
\end{align}
}
Along the $x^6$-direction, the M5-branes are placed at positions
\bea
0 \ \le \ a_1 \ \le \ a_2 \ \le \ \cdots \ \le \ a_N \ \le 2\pi R_6\,,
\eea
thereby partitioning the $x^6$ direction into $N$ intervals of length
{\allowdisplaybreaks
\begin{align}
t_{f_1} \ \ &=a_{2} - a_1\,, 
\nonumber\\
t_{f_2} \ \ &= a_3 - a_2 \, ,
\nonumber \\
& \quad \vdots \nonumber \\
t_{f_{N-1}} & = a_N - a_{N-1} \, ,
\nonumber \\
t_{f_N} \ & = 2 \pi R_6 - \sum_{i=1}^{N-1} t_{f_i} =2\pi R_6-(a_N-a_1) = -i \rho - (a_N - a_1)\,. \label{DistM5Branes}
\end{align}}
For a fixed $R_6$, the brane configuration is {specified} by $(N-1)$ independent non-negative {parameters}. The fugacity associated with these independent parameters $t_{f_i}, (i=1, \cdots, N-1)$ are denoted as 
\begin{align}
&Q_{f_1}=e^{-2 \pi t_{f_1}}\,, && Q_{f_2} = e^{-2 \pi t_{f_2}}\,, && \cdots  \,, && Q_{f_{N-1}} = e^{-2 \pi t_{f_{N-1}}}\,.\label{fugactitiesQf}
\end{align}
Thus, for meromorphic functions of the $Q_{f_i}$, we can view the complexified $ i t_{f_i}$ as $(N-1)$ independent positions on a torus $\mathbb{T}^2(\rho)$ of complex structure $\rho$.

\begin{figure} [htbp]
\centering
\vskip0cm
\makebox[\textwidth][c]{\includegraphics[width=1.0\textwidth]{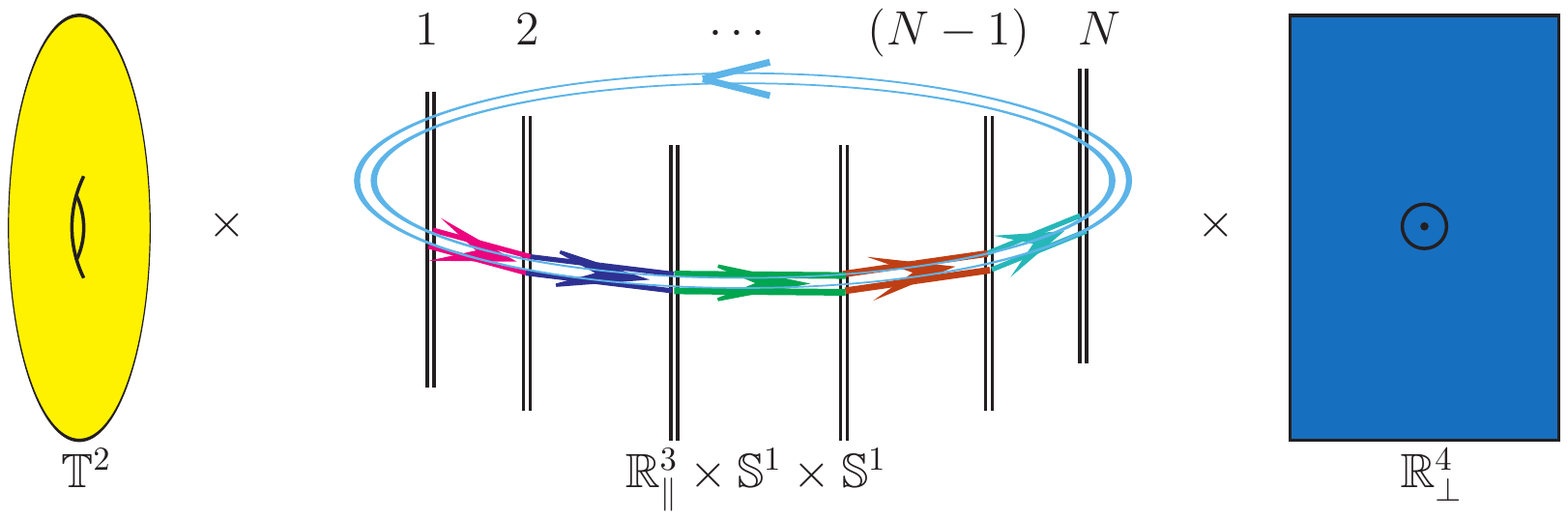}}
\vskip-16.5cm
\caption{\sl Brane configuration:
The M5-branes are all located at the origin in $\mathbb{R}^4_\perp$,
wrapped around $\mathbb{T}^2$ and stretched along the $(6)$-direction. }
\label{Fig:BraneConfig}
\end{figure}

The $K$ different M2-branes are stretched~\footnote{Since the transverse space $\mathbb{R}_\perp^4$ is topologically trivial, the M2-branes between any two M5-branes cannot be split but form a single stack (see \cite{Haghighat:2013tka,Hohenegger:2013ala}).} between the M5-branes and distributed among these $N$ intervals with multiplicities $(\{k_i\}) = (k_1, k_2, \cdots, k_N)$ such that $K = \sum_{i=1}^N k_i$. In addition, there are $M$ M-waves propagating along the intersections of M5- and M2-branes, \emph{i.e.} the directions $x^0$ and $x^1$. Finally, all branes are point-like and located at the origin with regards to $\mathbb{R}^4_\perp$.\footnote{We can also replace $\mathbb{R}^{4}_{\perp}$ by an affine $A_{N-1}$ geometry, which is dual to the M5-branes on a circle \cite{Anselmi:1993sm,Ooguri:1995wj,Kutasov:1995te}.} Schematically, the whole setup is shown in Figure~\ref{Fig:BraneConfig}.

The brane configuration saturates the BPS bound. Furthermore, the space-time Poincar\'e- and supersymmetry content is identical to that of the non-compact setting (\emph{i.e.} with $\mathbb{R}_{\parallel}^3\times\mathbb{S}^1_{R_6}$ replaced by $\mathbb{R}^4_{\parallel}$), which we already extensively discussed in \cite{Hohenegger:2015cba}. In section~\ref{Sect:PartitionFunction}, we present the partition function of this configuration. However, in order to render the latter well-defined, we need to regularize infrared divergences. To this end, we turn on various deformations of $\mathbb{R}^3_\parallel \times \mathbb{S}^1_{R_5}\times\mathbb{R}^4_\perp$, which can be described as a $U(1)_{\epsilon_1} \times U(1)_{\epsilon_2} \times U(1)_m$ action with respect to the $(0)$-direction. Specifically, for local coordinates $(z^1, z^2) = (x^2 + i x^3, x^4 + i x^5)$ and $(w^1, w^2) = (x^7 + i x^8, x^9 + i x^{10})$ of $\mathbb{R}^4_\parallel$, maximum deformations one can introduce with respect to $x^0$ are
\bea
U(1)_{\epsilon_1} \times U(1)_{\epsilon_2} \times U(1)_m: \quad
&& (z_1, \  z_2) \ \rightarrow \ \  (e^{2 \pi i \epsilon_1} z_1, e^{2 \pi i \epsilon_2} z_2) \nn
&& (w_1, w_2) \rightarrow \ \ (e^{2 \pi i m - i\pi (\epsilon_1 + \epsilon_2)} w_1, e^{ - 2 \pi i m - \pi i (\epsilon_1 + \epsilon_2)} w_2), \qquad\label{Deformations}
\eea
with the parameters $\epsilon_{1,2}$ and $m$. From the perspective of the four-dimensional ${\cal N}=2^*$ gauge theory, $\epsilon_{1,2}$ correspond to the deformation parameters of an $\Omega$-background \cite{Moore:1997dj,Lossev:1997bz,Nekrasov:2002qd} \footnote{Several different string theoretic descriptions of the $\Omega$-background have been proposed in the literature (see for example \cite{Billo:2006jm,Ito:2010vx,Huang:2010kf,Huang:2011qx,Hellerman:2011mv}.) In particular, a world-sheet approach based on physical scattering amplitudes has been proposed in \cite{Antoniadis:2010iq,Antoniadis:2013epe,Antoniadis:2013mna,Antoniadis:2015spa} (see also \cite{Nakayama:2011be}). Furthermore, in \cite{Bae:2015eoa} its relation to topological gravity has been understood.}, while $m$ can be associated with a mass deformation. In the present case, we are counting states on a partially compactified $\mathbb{R}^3_\parallel\times \mathbb{S}^1_{R_5}$. This space is not compatible with the above deformations. Therefore, in what follows, we shall take a suitable limit of the deformation that commutes with the isometries of $\mathbb{R}^3_\parallel \times \mathbb{S}^1_{R_5}$.

The $\Omega$ and mass deformations also affect the nature of three-torus $\mathbb{S}^1_{R_0} \times \mathbb{S}^1_{R_1} \times \mathbb{S}^1_{R_6}$. Among the three directions, the $x^0$-direction is twisted while the $x^1-$ and $x^6$-directions remain untwisted. So, we should expect for the deformed brane configuration that the full U-duality group of the brane configuration is reduced by the deformations but that the $\mathbb{Z}_2$ exchange symmetry between $\mathbb{S}^1_{R_1}$ and $\mathbb{S}^1_{R_6}$, {\sl i.e.} $\tau \leftrightarrow \rho$ in Eq.(\ref{tau-rho}), is still intact. 

Finally, we can also connect this configuration to a setup of D-branes in string theory: Indeed, by viewing $\mathbb{T}^2\sim \mathbb{S}^1\times \mathbb{S}^1$ (and particularly $x^0\sim x^0+2\pi R_0$), we can interpret the direction $x^0$ as the M-theory circle and dimensionally reduce to Type IIA string theory. In this way, the M5-branes are reduced to D4-branes, whose worldvolume dynamics is described by five-dimensional ${\cal N}=1^*$ gauge theory with coupling constant $g_5^2 = R_0$, the radius of the M-theory circle. The M2-branes become F1-strings with tension $T_2 R_1 R_6$, where $T_2$ is the M2-brane tension. 
\subsection{Calabi-Yau Geometry}\label{Sect:ToricCYXn}
\begin{figure}[h]
  \centering
  \includegraphics[width=4in]{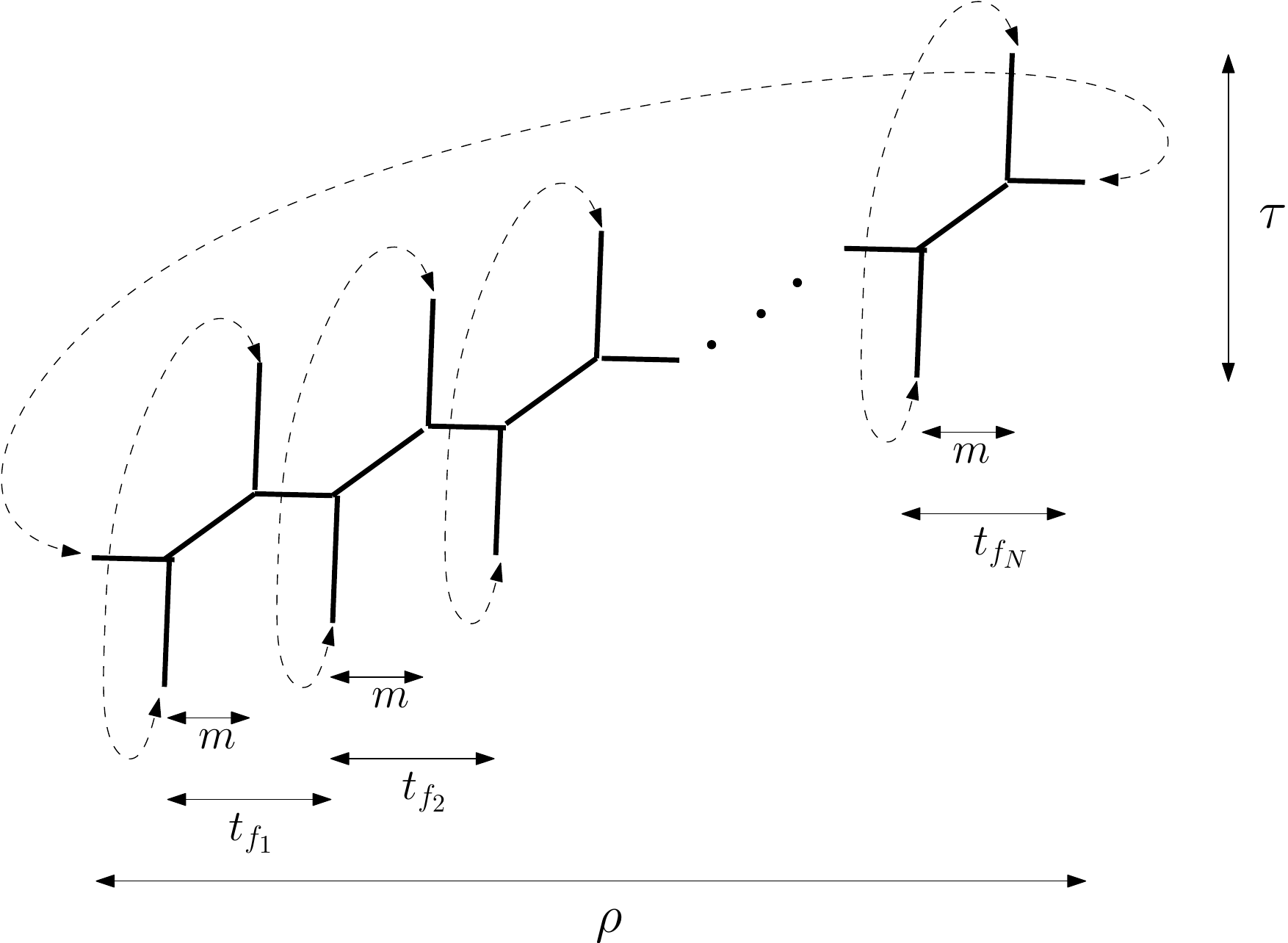}
  \caption{\sl The toric web diagram of the Calabi-Yau threefold $X_N$ dual to the brane configuration. Both the horizontal and vertical directions are compactified on $\mathbb{S}^1$s, defining the diagram on $\mathbb{T}^2$. 
  }\label{compacttoric}
\end{figure}

We can associate a toric Calabi-Yau threefold $X_N$ to the brane configuration just discussed,
whose web diagram is shown in Fig.~\ref{compacttoric}. In the toric diagram, the compactification of the vertical direction reflects the fact that the brane configuration is compactified along the $x^1$ direction, while the compactification of the horizontal direction reflects the fact that the brane configuration is compactified along the $x^6$ direction. Therefore, the toric web of $X_N$ is defined on a torus.

The new feature of this manifold in comparison to the non-compact configuration (\emph{i.e.} $R_6\to\infty$) discussed in \cite{Hohenegger:2015cba}, whose toric web is defined on a cylinder, is a two-fold fibration structure: the $X_N$ can be seen as an elliptic fibration over the affine ${A}_{N-1}$ space, which itself is an elliptic fibration over $\mathbb{C}^1$. Thus, $X_N$ is specified by three parameters, $\tau, \rho, m$, together with $N-1$ parameters appearing from resolution of affine $A_{N-1}$ singularities. The affine extension of $A_{N-1}$ is a direct consequence of compactifying $x^6\sim x^6+2\pi R_6$ in the brane setup. We will see below that this affine extension will play an important role in the gauge theory description.

This new structure can be made more transparent by using slightly different parameters than in the brane configuration. The latter is usually parametrized with the help of the distances {between} the M5-branes along $x^6$, \emph{i.e.} by $(\tau,m,t_{f_{1}},t_{f_{2}},\cdots, t_{f_{N}},\epsilon_1,\epsilon_2)$(see~(\ref{DistM5Branes})). We can replace one of these, \emph{i.e.} $t_{f_N}$, by the size of the circle transverse to the M5-branes
\begin{align}
\rho=i\sum_{a=1}^{N}t_{f_{a}}=2\pi i R_6\,,\label{Defrho}
\end{align}
and therefore use the parameters $(\tau,\rho,m,t_{f_{1}},t_{f_{2}},\cdots,t_{f_{N-1}},\epsilon_1,\epsilon_2)$ instead.

Recall that, in the toric web diagram Fig. \ref{compacttoric}, the presence of two $\mathbb{S}^1$s is associated with the two-fold fibration structure in $X_N$.  The exchange of these two $\mathbb{S}^1$s in the toric web amounts to an exchange of the { elliptic fiber and the elliptic base} in $X_N$. This implies that, in the M5-M2 brane configuration picture,  there is another configuration dual to the one discussed in~\ref{Sect:BraneConfig}: it is given by a single M5-brane wrapped on a circle with transverse space affine $A_{N-1}$ geometry and $N$ distinct M2-brane configurations. These two different brane configurations give rise to two dual gauge theory descriptions, as we shall discuss presently. \footnote{In the situation we have $A_{N-1}$ geometry rather than affine $A_{N-1}$ (i.e., $\rho \mapsto i\infty$), the two dual gauge theories were discussed in \cite{Hohenegger:2015cba}.}
\subsection{Gauge Theories with Affine Gauge Group}\label{Sect:GaugePerspective}
As explained in \cite{Hohenegger:2013ala}, we can associate $\Omega$- and mass-deformed supersymmetric gauge theories to the brane configuration discussed in section~\ref{Sect:BraneConfig}. In fact, the brane setup can be related by a chain of U-dualities to two distinct (but dual) gauge theories, which will play an important role throughout this paper:
\begin{itemize}
\item \textsc{ gauge theory 1}:\\
The first picture is to associate the K\"ahler parameter $\tau$ of the $\mathbb{T}^2$ with the coupling constant of an $U(N)$ gauge theory, while the K\"ahler parameters $t_{f_1}, \cdots, t_{f_N}$ are identified with the parameters of the Coulomb branch \footnote{It is known that, if the theory is coupled to $g$ many massless adjoint hypermultiplets, the partition function is equal to the partition function of a two-dimensional topological field theory on a genus-$g$ Riemann surface \cite{Iqbal:2015dra}.}. This theory is reduced to the $\mathcal{N}=2^*$ supersymmetric gauge theory in four dimensions.
\item \textsc{gauge theory 2}:\\
The second picture is to associate the K\"ahler parameters $t_{f_a}$'s of the base $\mathbb{P}^{1}$'s of $X_N$ with the coupling constants of a $[U(1)]^{N}$ quiver gauge theory. It is important to notice that, because the $x^6$ direction of the brane configuration is compactified on a circle (which gives rise to the affine $A_{N-1}$ structure of $X_N$), this quiver is circular rather than linear:
\begin{center}
\scalebox{1.2}{\parbox{9cm}{\begin{tikzpicture}
\draw[ultra thick] (0,0) circle (0.15cm);
\draw[ultra thick,-] (0.15,0) -- (0.85,0);
\draw[ultra thick] (1,0) circle (0.15cm);
\draw[ultra thick,-] (1.15,0) -- (1.85,0);
\draw[ultra thick] (2,0) circle (0.15cm);
\draw[ultra thick,-] (2.15,0) -- (2.6,0);
\node at (3,0) {$\cdots$};
\draw[ultra thick,-] (3.4,0) -- (3.85,0);
\draw[ultra thick] (4,0) circle (0.15cm);
\draw[ultra thick,-] (4.15,0) -- (4.85,0);
\draw[ultra thick] (5,0) circle (0.15cm);
\draw[ultra thick,-] (5.15,0) -- (5.85,0);
\draw[ultra thick] (6,0) circle (0.15cm);
\node at (-0.4,-0.5) {$U(1)_{1}$};
\node at (1.1,-0.5) {$U(1)_{2}$};
\node at (2.1,0.5) {$U(1)_{3}$};
\node at (4.1,0.5) {$U(1)_{N-3}$};
\node at (5.1,-0.5) {$U(1)_{N-2}$};
\node at (6.9,-0.5) {$U(1)_{N-1}$};
\draw[ultra thick] (3,2) circle (0.15cm);
\draw[ultra thick] (0.1,0.1) -- (2.9,1.9);
\draw[ultra thick] (5.9,0.1) -- (3.1,1.9);
\node at (3,2.5) {$U(1)_N$};
\end{tikzpicture}}}
\end{center}
\end{itemize}
For the reader's convenience, we compiled below the identification of all Calabi-Yau parameters with gauge theory parameters from the above two different pictures
\bea
\begin{array}{|c|c|c|c|c|}\hline
{\bf pm.} & \text{{\bf brane configuration}} & \text{\bf Calabi-Yau} & \text{{\bf gauge theory 1}} & \text{{\bf gauge theory 2}}\\\hline
&&&&\\[-8pt]
\tau&  \parbox{3.5cm}{size of $\mathbb{S}^1$ parallel \\
to M5-branes}  & \parbox{3.3cm}{K\"ahler moduli of \\ elliptic base} & \parbox{3.3cm}{coupling constant} & \parbox{3.5cm}{compact Coulomb \\ branch parameter}\\[12pt]\hline
&&&&\\[-8pt]
\rho & \parbox{3.5cm}{size of $\mathbb{S}^1$ transverse to M5 branes} &  \parbox{3.3cm}{K\"ahler moduli of \\ affine $ A_{N-1}$-fiber} & \parbox{3.5cm}{compact Coulomb \\ branch parameter} &  \parbox{3.3cm}{\quad \ \quad overall \\
coupling constant}  \\[12pt]\hline
&&&&\\[-8pt]
t_{f_{a}} & \parbox{4cm}{separations between \\
adjacent M5-branes}&\parbox{3cm} {\parbox{3.3cm}{K\"ahler moduli of \\
affine $ A_{N-1}$-fiber}}& \parbox{3.5cm}{compact Coulomb \\ branch parameter} &\parbox{3.3cm}{coupling constants \\ \quad $a=1,\ldots,N-1$}\\[12pt]\hline
\end{array}\nonumber
\eea
When counting the number of parameters, note that $\rho=\sum_{a=1}^{N}t_{f_a}$ and thus $(t_{f_1},\ldots, t_{f_N})$ and $\rho$ are not independent of one another. In all cases, $m$ and $\epsilon_{1,2}$ describe deformations.

\subsection{IIa and IIb Little String Theories}\label{Sect:LittleStrings}
The brane configuration discussed in section~\ref{Sect:BraneConfig} can also be related to little string theories, which are six-dimensional non-local quantum theories with non-gravitational string excitations \cite{Berkooz:1997cq,Seiberg:1997zk, Losev:1997hx, Aharony:1998ub}. We can associate type IIa and IIb little strings with Type IIB and IIA NS5-branes in the decoupling limit 
\begin{align}
&g_{\rm st} \rightarrow 0, &&  \ell_{\rm st} = \mbox{finite}
\label{decoupling}
\end{align}
for the string coupling and string length, respectively. %
At energies well below the string tension scale, the little string states are decoupled and the Type IIa and IIb little string theories flow to the $(1,1)$ super Yang-Mills theory and $(2, 0)$ superconformal theory, respectively. Notice, since the limit (\ref{decoupling}) commutes with T-duality (which  exchanges type IIA and IIB string theories), so the type IIa and IIb little string theories are also related by T-duality. We discuss the precise relation in section~\ref{TdualLittle}.

\subsubsection{Little String BPS Excitations}
We first explain how the little string theories are related to the M-brane configuration discussed in section~\ref{Sect:BraneConfig}. The BPS string excitations of little string theories are realized by the open M2-branes stretched between M5-branes. Since there are $N$ such intervals on the $\mathbb{S}^1$ transverse to $N$ M5-branes (\emph{i.e.} along the direction $x^6$), these excitations carry $[U(1)]^N$ quantum numbers whose chemical potentials and fugacities are $t_{f_1}, \cdots, t_{f_N}$, respectively, $(Q_{f_1}\,\ldots, Q_{f_N})$ in Eq.~(\ref{fugactitiesQf}). 

The crucial feature of the M-brane configuration that permits this identification with the the little string states is the compactness of the $x^6$ direction: Indeed, compared to the non-compact counterpart (as discussed in our previous paper \cite{Hohenegger:2015cba}), the parameters (\ref{DistM5Branes}) are modified in two important ways:
\begin{enumerate}
\item There is one additional finite interval between the first and the last M5-brane, which we denoted as $t_{f_N}$ in (\ref{DistM5Branes}). Therefore, even in the limit that all the $N$ M5-branes stack together and make the M-strings tensionless, there always exists a finite-tension string coming from the open M2-brane stretched around the compact $\mathbb{S}^1$ of the $x^6$-direction. This finite-tension string defines the little string. In our notation, the ground-state of a single little string corresponds to the configuration $(k_1, \cdots, k_N) = (1, \cdots, 1)$, \emph{i.e.} a closed M2-brane which pass through all $N$ M5-branes on $\mathbb{S}^1$. Likewise, the ground-state of $k$ multiple little strings corresponds to the configuration $(k_1, \cdots, k_N) = (k, \cdots, k)$, which can be multiply wound.  
\item The intervals $t_{f_1}, \cdots, t_{f_N}$ take values on a compact domain. More precisely, compared to the non-compact M-brane configuration, we have
\bea
0 \le t_{f_1}\leq \cdots\leq t_{f_N} < \infty \qquad \rightarrow \qquad 
0 \ \le \ t_{f_1}\leq \ t_{f_2} \leq \  \cdots \leq t_{f_{N}} \ \le \ 2\pi R_6 \ . 
\nonumber
\eea
This implies that the tensions of M-strings and little strings can only take a finite maximum value. This property is imperative for the little string theories to retain stringy features such as T-duality, as we discuss in the following subsubsection.
\end{enumerate}
\noindent
To explain better the nature of the little string BPS excitations, we can compare the multiple M5-branes on a transverse circle with multiple D$p$-branes on a transverse circle. In this comparison, we interpret the M-strings (\emph{i.e.} open M2-branes) as noncritical counterparts of open fundamental strings, while a little string ground state (defined by the configuration $(k_1, \cdots, k_N) = (1, \cdots, 1)$) is the noncritical counterpart of a closed fundamental string. This analogy points to two very important facts: First, in the same way as multiple open fundamental strings on the D$p$-branes can form a closed string and move freely in ambient ten-dimensional bulk spacetime, multiple open M2-branes ending on M5-branes can form a closed M2-brane and move freely in eleven-dimensional spacetime. Secondly, while the open fundamental strings can carry fractional winding number around the transverse circle, the M-strings also carry fractional winding numbers around the transverse circle. These are measured by the chemical potentials $(t_{f_1}, \cdots, t_{f_N})$ and the fugacities $(Q_{f_1}, \cdots, Q_{f_N})$. However, what makes the little strings very different from fundamental strings is that, in the decoupling limit Eq.(\ref{decoupling}), the little strings are confined inside the five-brane worldvolume, viz. the six-dimensional spacetime the little string theories live in.

\subsubsection{Relation to Gauge Theory and T-duality}\label{TdualLittle}
The above discussion establishes a connection between the little string theories and the M-brane configuration of section~\ref{Sect:BraneConfig}. Therefore, we can also relate the former to the two gauge theories that we discussed in section~\ref{Sect:GaugePerspective}. To make this connection precise, we first need to discuss the moduli spaces of type IIa and IIb little string theories and explain their connection to gauge theory 1 and gauge theory 2, respectively.

To this end, we begin in six-dimensions by first considering the direction $x^1$ in the M-brane configuration to be non-compact (\emph{i.e.} $R_1\to\infty$).\footnote{The direction $x^1$ is singled out since it is untwisted with respect to the deformations (\ref{Deformations}).}  In this framework, the non-chiral type IIa and the chiral IIb little string theories are defined on the six-dimensional worldvolume of the $N$ five-branes and preserve sixteen supercharges each. Their respective moduli spaces of supersymmetric vacua are
\begin{align}
&{\cal M}^{\rm 6d}_{\rm IIa} = {(\mathbb{R}^4)^N / S_N}\,, &&\text{and} &&{\cal M}^{\rm 6d}_{ \rm IIb} = {(\mathbb{R}^4 \times \mathbb{S}^1)^N / S_N}\,.\label{6dModSpaceLittle}
\end{align}
The $\mathbb{S}^1$ in ${\cal M}^{\rm 6d}_{\rm IIb}$ can be understood from the definition of the IIb little string theory in terms of the worldvolume of M5-branes. In the brane configuration of Section~\ref{Sect:BraneConfig}, it corresponds to the $\mathbb{S}^1_{R_6}$ of the compact $x^6$-direction. Notice that the two spaces (\ref{6dModSpaceLittle}) cannot be related to each other by any duality transformation. Indeed, from the perspective of the type IIA and IIB string theories, the only compact direction that is not twisted by (\ref{Deformations}) (and would therefore lend itself to T-duality) is $x^6$, which, however, is transverse to the five-branes. 

Next, we consider five-dimensional little string theories by taking the direction $x^1$ to be compact (\emph{i.e.} $R_1$ to be finite). This compactification has very different impacts on the two moduli spaces (\ref{6dModSpaceLittle}): On the one hand, the IIb moduli space remains the same, since the six-dimensional tensor multiplet does not generate a scalar when reduced to five dimensions. On the other hand, the moduli space of IIa little string theory gets enlarged, since the six-dimensional vector multiplet generates a scalar in five dimensions. This scalar comes from the Wilson loop around the dual circle $\widetilde{\mathbb S}^1_{1/R_1}$ and takes values over the interval $[0, R_1]$.\footnote{Here we are invoking that, starting from the compact M-brane configuration as defining IIb little string theory on $\mathbb{S}^1_{R_1}$,  compactification on the T-dual circle yields IIa little string theory on $\widetilde{\mathbb{S}}^1_{1/R_1}$.} Therefore, the moduli spaces of the five-dimensional little string theories are
\begin{align}
&{\cal M}^{\rm 5d}_ {\rm IIa} = {(\mathbb{R}^4 \times {\mathbb{S}}^1_{R_1})^N \over S_N}\,,&&\text{and} &&  
{\cal M}^{\rm 5d}_{\rm IIb} = {(\mathbb{R}^4 \times \mathbb{S}^1_{R_6})^N \over S_N}\,.
\label{modulispace}
\end{align}
We see that parameters of circle-compactified IIa and IIb little string theories are mapped to each other by the exchange of the radii 
\begin{align}
R_1\,\longleftrightarrow\, R_6\,,\label{Tduality}
\end{align}
while the parameters originating from $\mathbb{R}^4_\perp$ are the same. 

We stress that Eq.(\ref{Tduality}) is the manifestation of T-duality on the circle-compactified little string theories. 
Phrased differently, while from the perspective of the fundamental string theory the T-duality corresponds to the map $R_1\leftrightarrow 1/R_1$, from the perspective of the circle-compactified five-branes the T-duality manifests as exchanging circle-wrapped IIA and IIB five-branes. This T-duality commutes with the decoupling limit Eq.(\ref{decoupling}), so the T-duality on the circle-compactified IIa and IIb little string theories is realized by Eq.(\ref{Tduality}). 

Note also that, in the description in terms of the elliptically fibered Calabi-Yau manifold $X_N$, the exchange Eq.(\ref{Tduality}) corresponds to fiber-base duality, \emph{i.e.} the exchange of the two K\"ahler parameters $\tau$ and $\rho$ of $X_N$. 

With the moduli spaces identified for the circle-compactified little string theories, we are now ready to discuss their relation to the exact marginal couplings that specify the gauge theory descriptions introduced in section~\ref{Sect:GaugePerspective}. The U-duality map discussed in section \ref{Sect:BraneConfig} indicates that the IIa little string theory compactified on $\mathbb{S}^1_{R_1}$ is most naturally described by the Coulomb branch of the five-dimensional $U(N)$ gauge theory with the gauge coupling given by $\tau$. At a generic point of the Coulomb branch, the theory is described by a $[U(1)]^N$ quiver gauge theory, and therefore the Coulomb branch is spanned by $t_{f_1}, \cdots, t_{f_N}$. Thus, we identify \textsc{gauge theory 1} with the gauge theory description of the circle compactified IIa little string theory. 

Performing the T-duality $R_1 \rightarrow 1/R_1$, we obtain circle-compactified IIb little string theory, which is also described by a $[U(1)]^N$ quiver gauge theory. Since $\mathbb{S}^1_{R_1}$ spans part of the Coulomb branch (as becomes apparent from ${\cal M}^{\rm 5d}_{ \rm IIa}$ in (\ref{modulispace})), the gauge coupling constants must be encoded by the brane configuration along the $x^6$-direction. Indeed, they are given by $t_{f_1}, \cdots, t_{f_N}$, while $\tau$ is the Coulomb branch parameter. That is, we can identify \textsc{gauge theory 2} with the circle compactified IIb little string theory \footnote{Our identifications agree with the little string worldsheet description of \cite{Aharony:1999dw}, futher discussed in \cite{Kim:2015gha}.}.

The T-duality (\ref{Tduality}) between the five-dimensional IIa and IIb little string theories suggests that their partition functions $Z_{\rm IIa}$ and $Z_{\rm IIb}$ are related to each other upon exchange of $\tau$ and $\rho$
\begin{align}
Z_{\rm IIa}(\tau, \rho)=Z_{\rm IIb}(\rho,\tau)\,,\label{TdualPart}
\end{align}
where we have only displayed the dependence on $\tau$ and $\rho$ to save writing. Actually, the connection of the little string theories to the M-brane configuration discussed in section~\ref{Sect:BraneConfig} and the dual Calabi-Yau threefold $X_{N}$ suggests 
\begin{align}
&Z_{\rm IIa}(\tau, \rho) = {\cal Z}_{X_N} (\tau, \rho)\, &&\text{and} &&Z_{\rm IIb}(\tau, \rho) = {\cal Z}_{X_N} (\rho, \tau)\,,
\label{identification}
\end{align}
where ${\cal Z}_{X_N}(\tau,\rho,m,t_{f_{1}},\cdots,t_{f_{N-1}})$ is the topological string partition function associated with the elliptic Calabi-Yau threefold $X_N$.  This makes (\ref{TdualPart}) manifest. 

Indeed,
in section~\ref{Sect:MonopolesvsInst}, we provide relations between BPS counting functions of little string configurations with integer (\emph{i.e.} non-fractional) winding, respectively, momentum quantum numbers which are in line with this proposal. A more careful study of (\ref{identification}) for general configuration and its implications is currently under way \cite{ours}.




\section{Partition Functions}\label{Sect:PartitionFunction}
In this section, we obtain the partition function of BPS states corresponding to the brane configuration introduced in section~\ref{Sect:BraneConfig}. The most efficient way to compute the partition function is to begin from the geometric perspective, \emph{i.e.} with the toric Calabi-Yau threefold $X_N$ introduced in section~\ref{Sect:ToricCYXn}. The topological string partition function on $X_N$ will be denoted by $\mathcal{Z}_{X_N}(\tau,m,t_{f_{1}},\cdots,t_{f_{N}},\epsilon_1,\epsilon_2)$. It can subsequently be related to the partition function of the six-dimensional $\Omega$-deformed field theories discussed in section~\ref{Sect:GaugePerspective}.

\subsection{Topological String Partition Function}

The refined topological vertex formalism \cite{TV, Iqbal:2007ii} can be used to determine the topological string partition ${\cal Z}_{X_N}$ of toric Calabi-Yau threefold $X_{N}$ using its toric web diagram shown in \figref{compacttoric}. Recall the fact that $X_N$ can be related to two dual gauge theories (as discussed in section~\ref{Sect:GaugePerspective}) corresponds geometrically to fiber-base duality \cite{Katz:1997eq}. At a computational level, it is related to the choice of a "preferred direction" in the refined topological vertex formalism \cite{Iqbal:2007ii}. Specifically, we need to choose a set of parallel edges in the web in \figref{compacttoric} such that every vertex is one of the end points of one such edge. While the topological string partition function is independent of this choice (\emph{i.e.} it is the same for each such choice), it leads to different gauge theory interpretations of the partition function. From \figref{compacttoric}, it is clear that there are two distinct choices for the preferred direction: vertical or horizontal.

Before we discuss the form of the refined topological string partition function for a specific choice of the preferred direction,  let us recall that the refined topological string partition function captures the degeneracies of BPS states coming from 
 M2-branes wrapping the holomorphic curves in the Calabi-Yau threefold $X$ on which M-theory is compactified. If we
  denote by $N^{(j_{L},j_{R})}_{C}$ the number of BPS states, with spin content $(j_{L},j_{R})$ under the five dimensional little group $SU(2)_{L}\times SU(2)_{R}$, coming from an M2-brane wrapped the holomorphic curve $C$,  then the refined topological string partition function is given by \cite{Gopakumar:1998ii,Gopakumar:1998jq,Hollowood:2003cv}
 \bea\nonumber
 {\cal Z}_{X}&=&\mbox{PExp}(F_{X})\,,\\\nonumber
 F_{X}&=&\sum_{C\in H_{2}(X,\mathbb{Z})}e^{-A(C)}F_{C}(\epsilon_1,\epsilon_2)\,, \eea
 where $\mbox{PExp}$ is the plethystic exponential, $A(C)$ is the complexified area of $C$ and $F_{C}$ captures the degeneracies of single particle states coming from M2-branes wrapping $C\subset X$,
 \bea\nonumber
 F_{C}=\sum_{j_{L},j_{R}}N^{(j_{L},j_{R})}_{C}(-1)^{2j_{L}+2j_{R}}\Big[(\sqrt{\tfrac{t}{q}})^{-j_{R}}+\cdots (\sqrt{\tfrac{t}{q}})^{+j_{R}}\Big]\Big[(\sqrt{t\,q})^{-j_{L}}+\cdots +(\sqrt{t\,q})^{+j_{L}}\Big]\,,
 \eea
with $(q,t)=(e^{i\epsilon_{1}},e^{-i\epsilon_{2}})$. For generic Calabi-Yau threefold, $N^{(j_{L},j_{R})}$ can jump under complex structure deformations such that $\sum_{j_{R}}(-1)^{2j_{R}}N^{(j_{L},j_{R})}_{C}$ remains constant. Since toric Calabi-Yau threefolds do not admit any complex structure deformations therefore $N^{(j_{L},j_{R})}_{C}$ are topological invariants captured by the refined topological string partition function. In subsequent sections, we will consider $F_{C}$ for specific curve classes in the Calabi-Yau threefold $X_N$ and refer to it as the degeneracy counting function or just the counting function.

\subsubsection{Vertical Description}
 
If the preferred direction is chosen vertical, then the various partitions associated with the horizontal direction can be summed over completely to obtain $\mathcal{Z}_{X_{N}}(\tau,m,t_{f_{1}},\cdots,t_{f_{N}},\epsilon_1,\epsilon_2)$ (see \cite{Hohenegger:2013ala}):
\bea\label{GTPF1}
\mathcal{Z}_{X_{N}}(\tau,m,t_{f_{1}},\cdots,t_{f_{N}},\epsilon_{1,2})=Z_{1}(m,t_{f_{1}},\cdots,t_{f_{N}},\epsilon_{1,2})\,\widetilde{\cal Z}^{(1)}_{N}(\tau,m,t_{f_{1}},\cdots,t_{f_{N}},\epsilon_1,\epsilon_2)\,,
\eea
where $Z_{1}(m,t_{f_{1}},\cdots,t_{f_{N}},\epsilon_{1,2})$ is the part independent of $\tau$ and 
\bea\label{ExpandInstanton}
\widetilde{\cal Z}^{(1)}_{N}&=&\sum_{k\geq 0}Q_{\tau}^{k}\,C_{N,k}(m,t_{f_{1}},\cdots,t_{f_{N}},\epsilon_{1,2})\\
&=&\sum_{\alpha_{1}\cdots \alpha_{N}}Q_{\tau}^{|\alpha_{1}|+\cdots+|\alpha_{N}|}\,\prod_{a=1}^{N}\frac{\vartheta_{\alpha_{a}\alpha_{a}}(Q_{m})}{\vartheta_{\alpha_{a}\alpha_{a}}(\sqrt{\frac{t}{q}})}
\prod_{1\leq a<b\leq N}\frac{\vartheta_{\alpha_{a}\alpha_{b}}(Q_{ab}Q_{m}^{-1})\vartheta_{\alpha_{a}\alpha_{b}}(Q_{ab}Q_{m})}
{\vartheta_{\alpha_{a}\alpha_{b}}(Q_{ab}\sqrt{\frac{t}{q}})\vartheta_{\alpha_{a}\alpha_{b}}(Q_{ab}\sqrt{\frac{q}{t}})}\, \label{Z1part}
\eea
is the part that depends on $\tau$ through the fugacity $Q_\tau$. 
In (\ref{ExpandInstanton}), we denote integer partitions as $\alpha_{1}, \cdots, \alpha_N$. We also use the notation
\begin{align}
&Q_m=e^{2\pi im}\,, &&Q_\tau=e^{2\pi i\tau}\,, &&q=e^{i\epsilon_1}\,, &&t=e^{-i\epsilon_2}\,,&& Q_{ab}=\prod_{k=a}^{b-1}Q_{f_k}\, ,\label{NotationBasic1}
\end{align}
as well as
\bea
\vartheta_{\mu\nu}(x)&=&\prod_{(i,j)\in \mu}\theta_1(\rho;x^{-1}t^{-\nu^{t}_{j}+i-\frac{1}{2}}q^{-\mu_{i}+j-\frac{1}{2}})\prod_{(i,j)\in \nu}\theta_1(\rho;x^{-1}t^{\mu^{t}_{j}-i+\frac{1}{2}}q^{\nu_{i}-j+\frac{1}{2}})\,. 
\eea
Furthermore, $\theta_1(\tau;z)$ is one of the Jacobi theta functions (see \cite{Eichler} for further information)
\begin{align}
\theta_1(\rho;x)&=-iQ_\rho^{\frac{1}{8}}(x^{\frac{1}{2}}-x^{-\frac{1}{2}})\prod_{k=1}^{\infty}(1-Q_\rho^k)(1-x\,Q_{\rho}^{k})(1-x^{-1}Q_{\rho}^{k})\, . 
\end{align}
Recall that $\rho=2\pi i R_6$ (see eq.~(\ref{Defrho})) and $Q_\rho=e^{2\pi i \rho}$.

\noindent Associated with the partition function $\widetilde{{\cal Z}}^{(1)}_{N}$, we also consider the free energy
\begin{align}
\Sigma_N(\tau,\rho,m,t_{f_1},\ldots, t_{f_{N-1}},\epsilon_1,\epsilon_2)=\text{PLog}\,\widetilde{\mathcal{Z}}^{(1)}_N(\tau,\rho,m,t_{f_1},\ldots, t_{f_{N-1}},\epsilon_1,\epsilon_2)\,,
\end{align}
{defined in terms of} the plethystic logarithm of a function $f$
\begin{align}
\mbox{PLog}\, f(\omega,\epsilon_1,\epsilon_2):=\sum_{k=1}^{\infty}\frac{\mu(k)}{k}\mbox{ln}\,f(k\omega,k\epsilon_{1},k\epsilon_{2})\,,\label{PlethLog}
\end{align}
where $\mu(k)$ is the M\"obius function. Physically, the function $\Sigma_N$ counts single-particle BPS bound-states (see \cite{Sundborg:1999ue,Polyakov:2001af}). As in (\ref{ExpandInstanton}), we can equally introduce the fugacity expansion
\begin{align}
\Sigma_N(\tau,\rho,m,t_{f_1},\ldots, t_{f_{N-1}},\epsilon_1,\epsilon_2)=\sum_{k=0}^\infty Q_\tau^k \,\Sigma_{N,k}(\rho,m,t_{f_1},\ldots, t_{f_{N-1}},\epsilon_1,\epsilon_2)\,.
\end{align}
The coefficient functions can be further expanded in terms of the {$N-1$ relative K\"ahler parameter fugacities
$(Q_{f_1}, Q_{f_2}, \cdots, Q_{f_{N-1}})$:}
\begin{align}
\Sigma_{N,k}(\rho,m,t_{f_1},\ldots, t_{f_{N-1}},\epsilon_1,\epsilon_2)=\sum_{k_1,\ldots,k_{N-1}}Q_{f_1}^{k_1}\ldots Q_{f_{N-1}}^{k_{N-1}}\,\Sigma_{N,k}^{(k_1,\ldots,k_{N-1})}(\rho,m,\epsilon_1,\epsilon_2)\,.\label{SigmaDef}
\end{align}

\subsubsection{Horizontal Description} 
If in \figref{compacttoric} the preferred direction is chosen horizontal, then the topological string partition function $\mathcal{Z}_{X_{N}}(\tau,m,t_{f_{1}},\cdots,t_{f_{N}},\epsilon_1,\epsilon_2)$ has the form:
\bea\label{GTPF2}
\mathcal{Z}_{X_{N}}(\tau,m,t_{f_{1}},\cdots,t_{f_{N}},\epsilon_{1,2})=Z_{2}(N,\tau,m,\epsilon_{1,2})\,\widetilde{\mathcal{Z}}^{(2)}_{N}(\tau,m,t_{f_{1}},\cdots,t_{f_{N}},\epsilon_1,\epsilon_2)\,,
\eea
where $Z_{2}(N,\tau,m,\epsilon_{1,2})$ is the part independent of $t_{f_a}$. In order to write $\widetilde{\mathcal{Z}}^{(2)}_{N}(\tau,m,t_{f_{1}},\cdots,t_{f_{N}},\epsilon_1,\epsilon_2)$, 
we recall \cite{Haghighat:2013gba} that the topological string partition function can be obtained by gluing together building blocks $W_{\nu_{a}\nu_{a+1}}$ labelled by the partitions of integers $\nu_a$ and $\nu_{a+1}$. The $W_{\nu_a\nu_{a+1}}$ are open topological string amplitudes but can also be considered as capturing the BPS degeneracies of M2-branes ending on a single M5-brane from either side. The web diagram corresponding to this situation is shown in \figref{building} below.

\begin{figure}[h]
  \centering
  \includegraphics[width=1.6in]{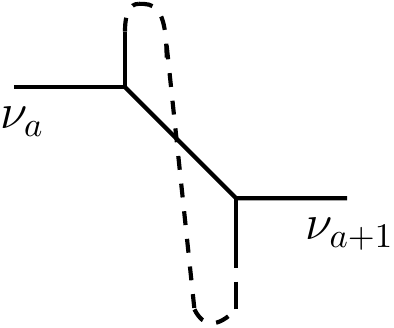}
 \caption{\sl The building block of partition function of configuration of M5-branes wrapping a circle.}\label{building}
\end{figure}

\noindent
The expression for $W_{\nu_{a}\nu_{a+1}}$ was calculated in \cite{Haghighat:2013gba} using the refined topological vertex formalism \cite{Hollowood:2003cv,Iqbal:2007ii} and is given by
\begin{align}
&W_{\nu_{a}^{t}\nu_{a+1}}(\tau,m,t,q)=W_{\emptyset\,\emptyset}(\tau,m,t,q)\,D_{\nu_{a}^{t}\nu_{a+1}}(\tau,m,t,q) \, , 
\label{buildingblock}
\end{align}
where
\bea
W_{\emptyset\,\emptyset}(\tau,m,t,q)=\prod_{k=1}^{\infty}\Big[(1-Q_{\tau}^{k})^{-1}\prod_{i,j=1}^{\infty}\frac{(1-Q_{\tau}^{k}Q_{m}^{-1}q^{-j+\frac{1}{2}}t^{-i+\frac{1}{2}})(1-Q_{\tau}^{k-1}Q_{m}q^{-j+\frac{1}{2}}t^{-i+\frac{1}{2}})}{(1-Q_{\tau}^{k}q^{-j+1}t^{-i})(1-Q_{\tau}^{k}q^{-j}t^{-i+1})}\,
\eea
and
\begin{align}
&D_{\nu_{a}^{t}\nu_{a+1}}(\tau,m,t,q)=\Big[t^{-\frac{\Arrowvert\nu^{t}_{a+1}\Arrowvert^2}{2}}\,q^{-\frac{\Arrowvert\nu_{a}\Arrowvert^2}{2}}Q_{m}^{-\frac{|\nu_a|+|\nu_{a+1}|}{2}}\Big]\nonumber\\
&\hspace{0.5cm}\times \prod_{k=1}^{\infty}\left[\prod_{(i,j)\in\nu_{a}}\frac{(1-Q_{\tau}^{k}Q_{m}^{-1}\,q^{-\nu_{a,i}+j-\frac{1}{2}}\,t^{-\nu_{a+1,j}^{t}+i-\frac{1}{2}})
(1-Q_{\tau}^{k-1}Q_{m}\,q^{\nu_{a,i}-j+\frac{1}{2}}\,t^{\nu_{a+1,j}^{t}-i+\frac{1}{2}})}{(1-Q_{\tau}^{k}\,q^{\nu_{a,i}-j}\, t^{\nu_{a,j}^{t}-i+1})(1-Q_{\tau}^{k-1}\,q^{-\nu_{a,i}+j-1}\,t^{-\nu_{a,j}^{t}+i})}\right. 
\nonumber\\
&\hspace{0.5cm}\left. \times \prod_{(i,j)\in\nu_{a+1}}\frac{(1-Q_{\tau}^{k}Q_{m}^{-1}\,q^{\nu_{a+1,i}-j+\frac{1}{2}}t^{\nu_{a,j}^{t}-i+\frac{1}{2}})(1-Q_{\tau}^{k-1}Q_{m} \,q^{-\nu_{a+1,i}+j-\frac{1}{2}}t^{-\nu_{a,j}^{t}+i-\frac{1}{2}})}{(1-Q_{\tau}^{k}\,q^{\nu_{a+1,i}-j+1}t^{\nu_{a+1,j}^{t}-i})(1-Q_{\tau}^{k-1}\,q^{-\nu_{a+1,i}+j}t^{-\nu_{a+1,j}^{t}+i-1})}\right]\, .
\end{align}
Here, our notation follows (\ref{NotationBasic1}). Furthermore, for a partition $\nu$ of length $\ell(\nu)$ we define 
\begin{align}
&|\nu|=\sum_{i=1}^{\ell(\nu)}\nu_i\,,&&||\nu||^2=\sum_{i=1}^{\ell(\nu)} \nu_i^2\, , 
\end{align}
and $\nu^t$ denotes the transposed partition. From (\ref{buildingblock}), the partition function can be calculated by gluing several $D_{\nu_{a}\nu_{a+1}}$ together by summing over the partitions $\nu_a$ and $\nu_{a+1}$. For example, the partition functions of $X_{2}$ which is dual to the brane configuration consisting of two M5-branes on the circle is given by
\begin{align}
{\cal Z}_{X_{2}}&=\sum_{\nu_1,\nu_2}(-Q_{f_{1}})^{|\nu_{1}|}(-Q_{f_{2}})^{|\nu_{2}|}\,W_{\nu_{1}\nu^{t}_{2}}(\tau,m,t,q)\,W_{\nu_{2}\nu^{t}_{1}}(\tau,m,q,t)\,.\label{DefX2}
\end{align}

For general $N$, the toric web diagram of the Calabi-Yau threefold $X_N$ that is dual to $N$ M5-branes distributed on $\mathbb{S}^1$-compactified $x^6$ direction 
is given in \figref{compacttoric}. The latter encodes how various $W_{\nu_a\nu_{a+1}}$ needs to be glued together to compute the partition function. Specifically,
\begin{align}
{\cal Z}_{X_{N}}(\tau,m,t_{f_{1}},\ldots,t_{f_N},\epsilon_1,\epsilon_2)&=&(W_{\emptyset\emptyset})^{N}\underbrace{\sum_{\nu_{1},\mathellipsis, \nu_{N}}\left(\prod_{a=1}^{N}(-Q_{f_{a}})^{|\nu_{a}|}\right)Z_{\nu_{1}\nu_{2}\cdots \nu_{N}}(\tau,m,\epsilon_1,\epsilon_2)}_{\widetilde{\mathcal{Z}}^{(2)}_{N}(\tau,m,t_{f_{1}},\cdots,t_{f_{N}},\epsilon_1,\epsilon_2)}\,,\label{DefPartFctComp0}
\end{align}
where the $t_{f_a}$-independent contribution in (\ref{GTPF2}) is given by
\begin{align}
Z_{2}(N,\tau,m,\epsilon_{1,2})=(W_{\emptyset\,\emptyset}(\tau,m,t,q))^N
\end{align}
and furthermore
\bea
Z_{\nu_{1}\nu_{2}\cdots \nu_{N}}=\left\{\begin{array}{ll}
D_{\nu_{1}\nu^{t}_{2}}(t,q)D_{\nu_{2}\nu^{t}_{3}}(q,t)
D_{\nu_{3}\nu^{t}_{4}}(t,q)\cdots D_{\nu^{t}_{N}\nu_{1}}(q,t) & \text{if }N \text{ is even,} \\[6pt]
D_{\nu_{1}\nu^{t}_{2}}(t,q)D_{\nu_{2}\nu^{t}_{3}}(q,t)
D_{\nu_{3}\nu^{t}_{4}}(t,q)\cdots D_{\nu^{t}_{N}\nu_{1}}(t,q) & \text{if }N \text{ is odd.} \end{array}\right.
\eea
Using Eq.(\ref{buildingblock}), the partition function can be written as
\begin{align}
\widetilde{\mathcal{Z}}^{(2)}_{N}(\tau,m,t_{f_{1}},\ldots,t_{f_N},\epsilon_1,\epsilon_2)&=\sum_{\nu_{1},\mathellipsis, \nu_{N}}\left(\prod_{a=1}^{N}(-Q_{f_{a}})^{|\nu_{a}|}\right)\,\prod_{a=1}^{N}\prod_{(i,j)\in \nu_{a}}\frac{\theta_{1}(\tau;z^{a}_{ij})\,\theta_{1}(\tau;v^{a}_{ij})}{\theta_1(\tau;w^{a}_{ij})\theta_1(\tau;u^{a}_{ij})}\,,\label{DefPartFctComp}
\end{align}
where the sum runs over the set of $N$ many integer partitions $\{\nu_{1},\nu_{2},\cdots, \nu_{N}\}$. For their arguments, we introduced the following short-hand notations
\bea \label{arguments}
z_{ij}^{a}&=&-m+\epsilon_{1}(\nu_{a,i}-j+\tfrac{1}{2})-\epsilon_{2}(\nu^{t}_{a+1,j}-i+\tfrac{1}{2})\,,\\\nonumber
v^{a}_{ij}&=&-m-\epsilon_{1}(\nu_{a,i}-j+\tfrac{1}{2})+\epsilon_{2}(\nu^{t}_{a-1,j}-i+\tfrac{1}{2})\,,\\\nonumber
w^{a}_{ij}&=&\epsilon_{1}(\nu_{a,i}-j+1)-\epsilon_{2}(\nu^{t}_{a,j}-i)\,,\\\nonumber
u^{a}_{ij}&=&\epsilon_{1}(\nu_{a,i}-j)-\epsilon_{2}(\nu^{t}_{a,j}-i+1)\,.
\eea
From the viewpoint of the brane configuration of section~\ref{Sect:BraneConfig}, the partition function (\ref{DefPartFctComp}) captures BPS excitations of the stretched M2-branes. The fact that M5-branes are distributed on $\mathbb{S}^1$-compactified $x^6$ direction is reflected in (\ref{DefPartFctComp}) through the identifications
\begin{align}
&\nu_{N+1}=\nu_{1}\,,&&\text{and} &&\nu_{0}=\nu_{N}\,.
\end{align}

\noindent Again, associated with the partition function
$\widetilde{\cal{Z}}_N^{(2)}$, we may introduce the free energy
\begin{align}
&\Omega_N(\tau,m,t_{f_{1}},\cdots,t_{f_{N}},\epsilon_1,\epsilon_2)=\mbox{PLog}\,\widetilde{\mathcal{Z}}^{(2)}_N(\tau,m,t_{f_{1}},\cdots,t_{f_{N}},\epsilon_1,\epsilon_2)\,,\label{FreeEnergies}
\end{align}
where PLog is defined in (\ref{PlethLog}). The free energy (\ref{FreeEnergies}) in turn can be expanded in powers of the K\"ahler moduli $(t_{f_1}, t_{f_2}, \cdots, t_{f_{N}})$ ( equivalently, $(t_{f_1}, t_{f_2}, \cdots, t_{f_{N-1}})$ and $\rho$):
\begin{align}
\Omega_N(\tau,m,t_{f_{1}},\cdots,t_{f_{N}},\epsilon_1,\epsilon_2)&=\sum_{k_1,\cdots,k_{N}\geq 0}Q_{f_{1}}^{k_{1}}\cdots Q_{f_{N}}^{k_{N}}\,G^{(k_1,\ldots,k_N)}(\tau,m,\epsilon_1,\epsilon_2)\, , \label{QExpansionMonopoles}
\end{align}
where $G^{(0,\cdots,0)} = 0$. Written in this form, the functions $G^{(k_{1},\ldots,k_{N})}$ encode the degeneracies of single-particle BPS bound-states in configurations with $N$ M5-branes distributed on a circle with $k_{i}$ M2-branes stretched between the $i$-th and the $(i+1)$-th M5-brane {for $i=1, \cdots, N$.} 

\subsubsection{Non-Compact Brane Configuration}
For completeness, we also present the topological string partition function for the case of a \emph{non-compact} $x^6$ direction, \emph{i.e.} for the case that the horizontal direction in  \figref{compacttoric} is decompactified to $\mathbb{R}^1$. From the brane configuration, this corresponds to the limit in which one of the distances $t_{f_a}$ is taken to infinity.

In the simplest case, for $N=2$, if we take the limit $Q_{f_{2}}\mapsto 0$ in (\ref{DefX2}), we get the partition function of the Calabi-Yau threefold $X_2$ which is an $A_{1}$ space fibered over $\mathbb{T}^2$ and is dual to the brane configuration in which we have two M5-branes on a line \emph{i.e.}, separated from each other by $t_{f_{1}}$,
\bea
{\cal Z}^{\text{line}}_{X_{2}}&=\sum_{\nu}(-Q_{f_{1}})^{|\nu|}\,W_{\emptyset\nu}(\tau,m,t,q)\,W_{\nu^{t}\emptyset}(\tau,m,q,t)\,.
\eea
More generally, the partition functions of $N\geq 2$ M5-brane separated along a non-compact direction $x^6$ can be obtained by restricting one of the partitions, say $\nu_N=\nu_0$, to be trivial
\begin{align}
\mathcal{Z}^{\text{line}}_{X_N}(\tau,m,t_{f_{1}},\ldots,t_{f_{N-1}},\epsilon_1,\epsilon_2)&=(W_{\emptyset\,\emptyset})^{N}\sum_{{\nu_{1},\mathellipsis, \nu_{N-1}}\atop{\nu_0=\nu_N=\emptyset}}\left(\prod_{a=1}^{N}(-Q_{f_{a}})^{|\nu_{a}|}\right)\,\prod_{a=1}^{N}\prod_{(i,j)\in \nu_{a}}\frac{\theta_{1}(\tau;z^{a}_{ij})\,\theta_{1}(\tau;v^{a}_{ij})}{\theta_1(\tau;w^{a}_{ij})\theta_1(\tau;u^{a}_{ij})}\,.\label{DefZline}
\end{align}
This is indeed the sole contribution to the partition function in the limit $Q_{f_N}=0$, corresponding to the infinite volume limit of $t_{f_N}$, which sends the interval between the first and $N$th M5-brane on $\mathbb{S}^1$ to infinity. The partition function $\mathcal{Z}_{X_N}^{\text{line}}$ has already been discussed in \cite{Haghighat:2013gba,Haghighat:2013tka,Hohenegger:2013ala,Hohenegger:2015cba}.

\subsection{Gauge Theory Partition Functions}
Given the topological string partition function $\mathcal{Z}_{X_N}(\tau,m,t_{f_{1}},\cdot, t_{f_{N}},\epsilon_{1,2})$, we can extract the instanton partition functions of the two gauge theories associated with $X_{N}$ as explained in section 2.3. This depends on the identification of the parameters of the affine $A_{N-1}$  fibration over $\mathbb{T}^2$ discussed earlier with the parameters of each gauge theory. 
\subsubsection{Gauge Theory 1}
We first discuss the reduction of the brane configuration (\ref{braneconfig}) over $\mathbb{S}^1(x^1)$ to a five-dimensional $U(N)$ gauge theory.  We identify the K\"ahler parameter $\tau$ of $\mathbb{T}^2$ with its gauge coupling constant, and extract the Nekrasov (instanton) partition function {by dividing out the classical and one-loop contribution} in the following manner
\bea\label{cnk}
\widetilde{\cal Z}^{(1)}_{N}(\tau,\rho, m,t_{f_{1}},\cdots,t_{f_{N-1}},\epsilon_1,\epsilon_2)&=&\frac{\mathcal{Z}_{X_N}(\tau,\rho,m,t_{f_{1}},\cdots,t_{f_{N-1}})}
{\lim_{\tau\mapsto i\infty}\mathcal{Z}_{X_N}(\tau,\rho,m,t_{f_{1}},\cdots,t_{f_{N-1}})}\,\\\label{PF}
&=&\sum_{k\geq 0}Q_{\tau}^{k}\,C_{N,k}(\rho,m,t_{f_{1}},\cdots,t_{f_{N-1}},\epsilon_1,\epsilon_2)\,,
\eea
Here, we have also identified the K\"ahler parameters of $X_N$ (which we parametrize by $t_{f_1},\ldots, t_{f_{N-1}}$ and $\rho$) with the gauge theory parameters of the configuration space $(\mathbb{S}^1)^N/S_N$. The explicit expression for $\widetilde{\mathcal Z}^{(1)}_{N}(\tau,m,t_{f_{1}},\cdots,t_{f_{N}},\epsilon_{1,2})$ is given in Eq.(\ref{Z1part}).
 
Thus, the topological string partition function $\mathcal{Z}_{X_{N}}$ is the supersymmetric partition function of {\sc gauge theory one} introduced in section~\ref{Sect:GaugePerspective}. the quantity $\widetilde{\cal Z}^{(1)}_{N}(\tau,\rho, m,t_{f_{1}},\cdots,t_{f_{N-1}},\epsilon_1,\epsilon_2)$ is its instanton contribution, \emph{i.e.} the coefficient function $C_{N,k}$ in (\ref{cnk}) encodes the charge $k$ instanton contribution. Specifically, if ${\cal M}(N,k)$ denotes the moduli space of $SU(N)$ instantons of charge $k$, then the coefficient $C_{N,k}$ is the elliptic genus of ${\cal M}(N,k)$ (see \cite{Hollowood:2003cv,Iqbal:2007ii}):
\bea
C_{N,k}(m,t_{f_1},\ldots,t_{f_{N}},\epsilon_1,\epsilon_2)&=\phi_{{\cal M}(N,k)}(\rho,m,t_{f_1},\ldots,t_{f_{N-1}},\epsilon_1,\epsilon_2)\,,
\eea
with $\rho$ being the elliptic parameter of the elliptic genus. Furthermore, $(t_{f_{1}},\cdots,t_{f_{N-1}})$ are the equivariant deformation parameters associated with the Cartan $U(1)^{N-1}$ global symmetry and $(\epsilon_1,\epsilon_2)$ are the equivariant parameters of the $U(1)\times U(1)$ action on ${\cal M}(N,k)$ coming from the Cartan of the $SO(4)$ action on $\mathbb{C}^2$.

Finally, in light of the discussion in Section~\ref{Sect:LittleStrings}, we see that the coefficients $\Sigma_{N,k}^{(k_1,\ldots,k_{N-1})}(\rho,m,\epsilon_1,\epsilon_2)$ defined in (\ref{SigmaDef}) encode the BPS degeneracies of type IIa little strings with charge configuration $(k_1\cdots,k_{N-1})$. 
 
\subsubsection{Gauge Theory 2}
Upon T-dualizing along $\mathbb{S}^1(x^6)$, the M5-branes are mapped to an affine $A_{N-1}$ geometry. This gives a five-dimensional $[U(1)]^N$ affine quiver gauge theory. We identify the K\"ahler parameters $t_{f_{1}}, t_{f_2}, \cdots, t_{f_N}$, equivalently, $(t_{f_1}, t_{f_2}, \cdots, t_{f_{N-1}})$ and $\rho$ with the gauge coupling constants, and extract BPS state partition function by dividing out vacuum contribution
\bea\nonumber
\widetilde{{\cal Z}}^{(2)}_{N}(\tau,m,t_{f_{1}},\cdots,t_{f_{N}},\epsilon_1,\epsilon_2)&=&\frac{\mathcal{Z}_{X_N}(\tau,m,t_{f_{1}},\cdots,t_{f_{N}})}
{\lim_{Q_{f_{1}}\mapsto 0}\ldots \lim_{Q_{f_{N}}\mapsto 0}\mathcal{Z}_{X_N}(\tau,m,t_{f_{1}},\cdots,t_{f_{N}},\epsilon_1,\epsilon_2)}\,,\\
&& \nonumber \\
&=&\sum_{\nu_{1},\mathellipsis, \nu_{N}}\left(\prod_{a=1}^{N}(-Q_{f_{a}})^{|\nu_{a}|}\right)Z_{\nu_{1}\nu_{2}\cdots \nu_{N}}(\tau,m,\epsilon_1,\epsilon_2)
\eea
The explicit form of $\widetilde{{\cal Z}}^{(2)}_{N}$ is already given in Eq.(\ref{DefPartFctComp}), so the coefficient functions are 
\begin{align}
Z_{\nu_{1}\nu_{2}\cdots \nu_{N}}(\tau,m,\epsilon_1,\epsilon_2)=\prod_{a=1}^{N}\prod_{(i,j)\in \nu_{a}}\frac{\theta_{1}(\tau;z^{a}_{ij})\,\theta_{1}(\tau;v^{a}_{ij})}{\theta_1(\tau;w^{a}_{ij})\theta_1(\tau;u^{a}_{ij})} \, . 
\end{align}
Thus, the topological string partition function ${\cal Z}_{X_N}$ 
is the supersymmetric partition function of the \textsc{gauge theory two} of section 2.3, and the corresponding 
$\widetilde{\cal Z}^{(2)}_N$ contains the contribution of BPS excitations. Since the gauge theory is $[U(1)]^N$ quiver gauge theory therefore the point like instantons are labelled by $(k_1,k_2,\cdots,k_N)$ where $k_a$ is the point like instanton charge for the $a$-th factor. The corresponding instanton moduli space is $N_{k_1,\cdots,k_N}:=\mbox{Hilb}^{k_{1}}[\mathbb{C}^2]\times \mbox{Hilb}^{k_{2}}[\mathbb{C}^2]\times \cdots \times \mbox{Hilb}^{k_{N}}[\mathbb{C}^2]$  where $\mbox{Hilb}^{k}[\mathbb{C}^2]$ is the Hilbert scheme of $k$ points on $\mathbb{C}^2$. The coefficient functions $Z_{k_{1}\cdots k_{N}}$ are given by an equivariant integral over $N_{k_1,\cdots, k_N}$  \cite{Haghighat:2013gba, Hohenegger:2013ala}.

In light of the discussion in Section~\ref{Sect:LittleStrings}, we see that the coefficients $G^{(k_1,\ldots,k_N)}(\tau,m,\epsilon_1,\epsilon_2)$ encode degeneracies of type IIb little strings with charge configuration $(k_1, \cdots, k_N)$.

\subsubsection{Non-Compact Partition Function}
{For comparison, we also recall} the instanton partition function in the limit $R_6\to\infty$
\begin{align}\nonumber
\widetilde{{\cal Z}}^{\text{line}}_{N}(\tau,m,t_{f_{1}},\cdots,t_{f_{N-1}},\epsilon_1,\epsilon_2)&=\frac{\mathcal{Z}^{\text{line}}_{X_N}(\tau,m,t_{f_{1}},\cdots,t_{f_{N-1}})}
{\lim_{Q_{f_{1}}\mapsto 0}\ldots \lim_{Q_{f_{N-1}}\mapsto 0}\mathcal{Z}^{\text{line}}_{X_N}(\tau,m,t_{f_{1}},\cdots,t_{f_{N-1}},\epsilon_1,\epsilon_2)}\,,\\\nonumber
&=\sum_{{\nu_{1},\mathellipsis, \nu_{N-1}}\atop{\nu_0=\nu_N=\emptyset}}\left(\prod_{a=1}^{N}(-Q_{f_{a}})^{|\nu_{a}|}\right)\,\prod_{a=1}^{N}\prod_{(i,j)\in \nu_{a}}\frac{\theta_{1}(\tau;z^{a}_{ij})\,\theta_{1}(\tau;v^{a}_{ij})}{\theta_1(\tau;w^{a}_{ij})\theta_1(\tau;u^{a}_{ij})}\,.
\end{align}
where $\mathcal{Z}^{\text{line}}_{X_N}$ is introduced in {(\ref{DefZline}).} We can similarly define the free energy
\begin{align}
\Omega^{\text{line}}_N(\tau,m,t_{f_1},\ldots,t_{f_{N-1}},\epsilon_{1},\epsilon_{2})=\mbox{PLog}\,\widetilde{\mathcal{Z}}^{\text{line}}_N(\tau,m,t_{f_1},\ldots,t_{f_{N-1}},\epsilon_{1},\epsilon_{2})\, , 
\end{align}
which we can expand in counting functions of single-particle BPS bound-states
\begin{align}
\Omega^{\text{line}}_N(\tau,m,t_{f_1},\ldots,t_{f_{N-1}},\epsilon_{1},\epsilon_{2})&=\sum_{k_1,\cdots,k_{N-1}\geq 0}Q_{f_{1}}^{k_{1}}\cdots Q_{f_{N-1}}^{k_{N-1}}\,F^{(k_1,\ldots,k_{N-1})}(\tau,m,\epsilon_1,\epsilon_2)\,.
\end{align}
We discussed the properties of $\widetilde{\mathcal{Z}}^{\text{line}}_{N}$ and $F^{(k_{1},\ldots,k_{N-1})}$ in great detail in \cite{Hohenegger:2015cba}. The latter counts the BPS bound-states of configurations in which $N$ M5-branes are distributed along a non-compact direction with $k_{i}$ M2-branes stretched between the $i$-th and the $(i+1)$-th M5-brane.

For the reader's convenience, we provide the following overview of the notation for the three different theories
\begin{center}
\begin{tabular}{|c|c|c|c|}\hline
{\bf quantity} & {\bf gauge theory 1} & {\bf gauge theory 2} & {\bf non-compact theory} \\ \hline
&&&\\[-12pt]
variables & $\tau,\rho,m,t_{f_1},\ldots t_{f_{N-1}},\epsilon_{1,2}$ & $\tau,m,t_{f_1},\ldots t_{f_{N}},\epsilon_{1,2}$ & $\tau,m,t_{f_1},\ldots t_{f_{N-1}},\epsilon_{1,2}$ \\[4pt]\hline
&&&\\[-8pt]
partition
function & $\widetilde{\mathcal{Z}}^{(1)}_N(\tau,\rho,m,t_{f_a},\epsilon_{1,2})$ & $\widetilde{\mathcal{Z}}^{(2)}_N(\tau,m,t_{f_a},\epsilon_{1,2})$ & $\widetilde{\mathcal{Z}}^{\text{line}}_N(\tau,m,t_{f_a},\epsilon_{1,2})$\\[4pt]\hline
&&&\\[-12pt]
free energy & $\Sigma_{N,k}(\rho,m,t_{f_a},\epsilon_{1,2})$ & $\Omega_N(\tau,m,t_{f_a},\epsilon_{1,2})$ & $\Omega^{\text{line}}_N(\tau,m,t_{f_a},\epsilon_{1,2})$\\[4pt]\hline
&&&\\[-12pt]
counting
functions & $\Sigma^{(\{k_i\})}_{N,k}(\rho,m,\epsilon_{1,2})$ & $G^{(\{k_i\})}(\tau,m,\epsilon_{1,2})$ & $F^{(\{k_i\})}(\tau,m,\epsilon_{1,2})$\\[4pt]\hline
\end{tabular}
\end{center}

In \cite{Hohenegger:2015cba} it was argued that $\lim_{\epsilon_2\to 0}F^{(k_{1},\cdots,k_{N-1})}(\tau,m,\epsilon_1,\epsilon_2)$ are related to the equivariant elliptic genus of the moduli space of monopole strings with charge $(k_1,k_2,\cdots,k_{N-1})$. More precisely, if ${\cal M}_{k_1,\cdots,k_{N-1}}$ is the moduli space of charge $(k_1,\cdots,k_{N-1})$ monopoles then its elliptic genus $\phi({\cal M}_{k_1,\cdots,k_{N-1}})$ is given by
\bea\label{onenewequation}
\phi({\cal M}_{k_1,\cdots,k_{N-1}})=\lim_{\epsilon_2\mapsto 0}\,\frac{F^{(k_1,\cdots,k_{N-1})}(\tau,m,\epsilon_1,\epsilon_2)}{F^{(1)}(\tau,m,\epsilon_1,\epsilon_2)}
\eea
Let us define the analog of the right hand side of the above equation for the compact brane configuration case,
\bea
P_{k_1,\cdots,k_N}(\tau,m,\epsilon_1):=\lim_{\epsilon_2\mapsto 0}\,\frac{G^{(k_1,\cdots,k_{N})}(\tau,m,\epsilon_1,\epsilon_2)}{G^{(1)}(\tau,m,\epsilon_1,\epsilon_2)}\,.
\eea
The function $P_{k_1,\cdots,k_N}(\tau,m,\epsilon_1)$ have modular properties very similar to right hand side of Eq.(\ref{onenewequation}),
\bea
P_{k_1,\cdots,k_N}(\tau+1,m,\epsilon_1)&=&P_{k_1,\cdots,k_N}(\tau,m,\epsilon_1)\\\nonumber
P_{k_1,\cdots,k_N}(-\frac{1}{\tau},\frac{m}{\tau},\frac{\epsilon_1}{\tau})&=&e^{\frac{2\pi i(m^2-\epsilon_{1}^2)}{\tau}(K-1)}P_{k_1,\cdots,k_N}(\tau,m,\epsilon_1)\,,\\\nonumber
P_{k_1,\cdots,k_N}(\tau,m+\ell \tau+r,\epsilon_1)&=&e^{-2\pi i K\ell^2\tau+4\pi i m K}P_{k_1,\cdots,k_N}(\tau,m,\epsilon_1)\,,
\eea
where $K=k_{1}+\cdots+k_N$. These modular transformation properties together with relation between $F^{(k_1,\cdots,k_{N-1} )}$ and $G^{(k_1,\cdots,k_{N-1})}$ leads us to conjecture that $P_{k_1,\cdots,k_N}(\tau,m,\epsilon_1)$ is the equivariant elliptic genus of the moduli space of monopoles of charge $(k_1,\cdots,k_N)$.  More specifically, if we denote the relative moduli space of affine $A_{N-1}$ monopoles of charge $(k_1,\cdots,k_N)$ by ${\cal M}^{KK}_{k_1,\cdots,k_N}$ then,
\bea
\phi({\cal M}^{KK}_{k_1,\cdots,k_{N}})=P_{k_1,\cdots,k_N}(\tau,m,\epsilon_1).
\eea


\subsection{Modular Properties}
The topological string partition function $\mathcal{Z}_{X_N}$ depends on two different modular parameters, $\tau$ and $\rho$. These transform under the $SL(2,\mathbb{Z})\times SL(2,\mathbb{Z})$ modular group action in the following manner:\footnote{Here, we choose a convention in which we treat $(t_{f_1},\ldots,t_{f_{N-1}},\rho)$ as independent variables.}
\begin{align}\label{ModTrans2}
(\tau,\rho,m,t_{f_{a}},\epsilon_1,\epsilon_2)&\mapsto\Big(\frac{a\tau+b}{c\tau+d},\rho, \frac{m}{c\tau+d}, t_{f_{a}},\frac{\epsilon_1}{c\tau+d},\frac{\epsilon_2}{c\tau+d}\Big)\,,\\\label{ModTrans1}
(\tau,\rho,m,t_{f_{a}},\epsilon_1,\epsilon_2)&\mapsto \Big(\tau, \frac{a\rho+b}{c\rho+d},\frac{m}{c\rho+d},\frac{t_{f_{a}}}{c\rho+d},\frac{\epsilon_1}{c\rho+d},\frac{\epsilon_2}{c\rho+d}\Big)\,,
\end{align}
where $\begin{bmatrix}
a & b \\
c & d
\end{bmatrix}\in SL(2,\mathbb{Z})$. The Calabi-Yau threefold $X_{N}$ is an affine $A_{N-1}$ space fibered over $\mathbb{T}^2$. In this geometric description, $\tau$ is the K\"ahler parameter of the base and therefore the fiber parameters are neutral under the modular transformation (\ref{ModTrans2}) (see \cite{Lockhart:2012vp}). The parameter $\rho$ is the K\"ahler parameter of the elliptic fiber in the affine $A_{N-1}$ space.

We will see that the topological string partition function $\mathcal{Z}_{X_{N}}(\tau,\rho,m,t_{f_1},\ldots,t_{f_{N-1}},\epsilon_1,\epsilon_2)$ is invariant (modulo a holomorphic anomaly \cite{Haghighat:2013gba,Haghighat:2013tka,Hohenegger:2013ala,Hohenegger:2015cba} and non-perturbative corrections  \cite{Lockhart:2012vp}) under the above transformations \emph{i.e.}$\, \mathcal{Z}_{X_{N}}$ is manifestly invariant under $SL(2,\mathbb{Z})\times SL(2,\mathbb{Z})$ modular group action. The full invariance group might actually be larger, as in the case $N=1$ for which the full invariance group is $Sp(2,\mathbb{Z})$ \cite{Hollowood:2003cv}.

\subsubsection{Transformation $\tau \mapsto\frac{a\tau+b}{c\tau+d}$}\label{SS1}

To show that $\mathcal{Z}_{X_{N}}$ is invariant under Eq.(\ref{ModTrans2}), we use its form given by Eq.(\ref{DefPartFctComp0}),
\bea
\mathcal{Z}_{X_{N}}(\tau,m,t_{f_{1}},\cdots,t_{f_{N}},\epsilon_{1,2})=(W_{\emptyset\emptyset})^{N}\,\widetilde{\mathcal{Z}}^{(2)}_{N}(\tau,m,t_{f_{1}},\cdots,t_{f_{N}},\epsilon_{1},\epsilon_{2})\,.
\eea
The function $\widetilde{\mathcal{Z}}^{(2)}_{N}(\tau,m,t_{f_{1}},\cdots,t_{f_{N}},\epsilon_{1},\epsilon_{2})$ is a sum of a product of Jacobi theta functions $\theta_{1}(\tau,z)$ given by
\bea\label{XP1}
\widetilde{\mathcal{Z}}^{(2)}_{N}(\tau,m,t_{f_{1}},\cdots,t_{f_{N}},\epsilon_{1},\epsilon_{2})=\sum_{\nu_{1}\cdots\nu_{N}}\prod_{a=1}^{N}(-Q_{f_{a}})^{|\nu_{a}|}\prod_{(i,j)\in \nu_{a}}\frac{\theta_{1}(\tau;z^{a}_{ij})\,\theta_{1}(\tau;v^{a}_{ij})}{\theta_1(\tau;w^{a}_{ij})\theta_1(\tau;u^{a}_{ij})}\,,
\eea
where $z_{ij}^{a},v_{ij}^{a},w_{ij}^{a}$ and $u_{ij}^{a}$ are given in Eq.(\ref{arguments}). The $\theta_{1}(\tau,z)$ transform under $\tau\mapsto -\frac{1}{\tau}$ in the following manner
\bea
\frac{\theta_{1}(-\frac{1}{\tau},\frac{z_{1}}{\tau})}{\theta_{1}(-\frac{1}{\tau},\frac{z_{2}}{\tau})}=e^{\frac{i\pi}{\tau}(z_{1}^{2}-z_{2}^2)}\,\frac{\theta_{1}(\tau,z_{1})}{\theta_{1}(\tau,z_{2})}\,.
\eea
To understand the nontrivial phase factor, we recall that $\theta_{1}(\tau,z)$ can be expressed as
\begin{align}
\theta_1(\tau,z)=\eta^3(\tau)\,(2\pi iz)\,\text{exp}\left[\sum_{k=1}^\infty \frac{B_{2k}}{(2k)(2k)!}\,E_{2k}(\tau)\,(2\pi iz)^{2k}\right]\,,\label{ExprTheta1}
\end{align}
where $\eta(\tau)$ is the Dedekind eta-function, $B_{2k}$ are the Bernoulli numbers and $E_{2k}(\tau)$ are the Eisenstein series. Eq.(\ref{ExprTheta1}) in particular also contains $E_{2}(\tau)$, which is holomorphic but not a modular form. It is well known that by adding a term 
\begin{align}
E_{2}(\tau)\mapsto \hat{E}_2(\tau,\bar{\tau})=E_2(\tau)-\frac{3}{\pi\,\text{Im}\tau}\,,
\end{align}
it can be made into a modular form of weight $2$. However, since the added term is not holomorphic in $\tau$, $\hat{E}_2(\tau,\bar{\tau})$ is non holomorphic. If we introduce
\begin{align}
\hat{\theta}_1(\tau,z)=\eta^3(\tau)\,(2\pi iz)\,\text{exp}\left[\frac{(2\pi iz)^2}{24}\,\hat{E}_2(\tau,\bar{\tau})+\sum_{k=2}^\infty \frac{B_{2k}}{(2k)(2k)!}\,E_{2k}(\tau)\,(2\pi iz)^{2k}\right]\,,\label{ExprTheta2}
\end{align}
then the replacement
\bea
\prod_{a=1}^{N}\prod_{(i,j)\in \nu_{a}}\frac{\theta_{1}(\tau;z^{a}_{ij})\,\theta_{1}(\tau;v^{a}_{ij})}{\theta_1(\tau;w^{a}_{ij})\theta_1(\tau;u^{a}_{ij})}\ \mapsto \prod_{a=1}^{N}\prod_{(i,j)\in \nu_{a}}\frac{\widehat{\theta}_{1}(\tau;z^{a}_{ij})\,\widehat{\theta}_{1}(\tau;v^{a}_{ij})}{\widehat{\theta}_1(\tau;w^{a}_{ij})\widehat{\theta}_1(\tau;u^{a}_{ij})}
\eea
in Eq.(\ref{XP1}) makes $\widetilde{\mathcal{Z}}^{(2)}_{N}(\tau,m,t_{f_{1}},\cdots,t_{f_{N}},\epsilon_{1},\epsilon_{2})$ modular invariant under Eq.(\ref{ModTrans2}). Similarly, since $W_{\emptyset\emptyset}$ is a ratio of products of $\theta_{1}(\tau,z)$, it too becomes modular invariant under similar replacement \footnote{Since it is a product of infinite number of theta functions its modular properties are better understood by writing it in terms of double elliptic Gamma function. In this way, one can show that it satisfies a non-perturbative modular transformation, \emph{i.e.} that it is modular invariant up to non-perturbative corrections in $\Omega$-deformation parameters 
\cite{Lockhart:2012vp}.}. Thus the complete partition function $\mathcal{Z}_{X_{N}}$ is invariant under modular transformation modulo holomorphic anomaly, introduced by $\theta(\tau,z)\mapsto \hat{\theta}_{1}(\tau,z)$, and possible non-perturbative corrections.

This is a good place to contrast the compact situation we presently consider with the non-compact situation. Without the replacement $\theta(\tau,z)\mapsto \hat{\theta}_{1}(\tau,z)$, the summand in Eq.(\ref{XP1}) 
\begin{align}
Z_{\nu_{1}\cdots\nu_{N}}(\tau,m,\epsilon_1,\epsilon_2)&=\prod_{a=1}^{N}\prod_{(i,j)\in \nu_{a}}\frac{\theta_{1}(\tau;z^{a}_{ij})\,\theta_{1}(\tau;v^{a}_{ij})}{\theta_1(\tau;w^{a}_{ij})\theta_1(\tau;u^{a}_{ij})}\,
\end{align}
transforms by a phase factor
\bea
Z_{\nu_1,\ldots,\nu_N}\left(-\tfrac{1}{\tau},\tfrac{m}{\tau},\tfrac{\epsilon_1}{\tau},\tfrac{\epsilon_2}{\tau}\right)= e^{\frac{2\pi i r_{\vec{\nu}}}{\tau}}\,Z_{\nu_1,\ldots, \nu_N}(\tau,m,\epsilon_1,\epsilon_2)\,.
\eea
Here,
\bea
r_{\vec{\nu}}(m,\epsilon_1,\epsilon_2)=\frac{1}{2}\sum_{a=1}^{N}\sum_{(i,j)\in \nu_a}\Big((z^{a}_{ij})^2+(v^{a}_{ij})^2-(w^{a}_{ij})^2-(u^{a}_{ij})^2\Big)\,,
\eea
which depends explicitly on the {shape} of the partitions $\{\nu_1,\ldots,\nu_N\}$. This is in contrast to the non-compact case $\mathcal{Z}_N^{\text{line}}$ in Eq.(\ref{DefZline}): as discussed in \cite{Hohenegger:2015cba}, for $\nu_N=\emptyset$, $r_{\nu_1,\ldots,\nu_{n-1},\emptyset}(m,\epsilon_1,\epsilon_2)$ depends only on the size of the partitions $|\nu_1|,\ldots,|\nu_{N-1}|$ but not on their shape. Roughly speaking, this difference between non-compact and compact situations originates from whether the endpoint partitions $\nu_0, \nu_N$ are trivial or not. 

For non-trivial $\nu_{N}$, we can write $r_{\vec{\nu}}$ in the following suggestive form
\begin{align}
&r_{\vec{\nu}}(m,\epsilon_1,\epsilon_2)=K\,m^2+(p_{\vec{\nu}}-\tfrac{K}{2})\epsilon_{+}^2+(-p_{\vec{\nu}}-\tfrac{K}{2})\epsilon_{-}^2\, &&\text{with} &&K=\sum_{i=1}^N|\nu_i|\,,\label{phase}
\end{align}
where only $p_{\vec{\nu}}$ depends on the form of the partitions. From the brane configuration point of view, $K$ corresponds to the total number of M2-branes stretched between the $N$ M5-branes. It is clear from Eq.(\ref{phase}) that the partition function (\ref{DefPartFctComp}) has interesting modular properties in the Nekrasov-Shatashvili (NS) limit $\epsilon_2\to 0$ (see \cite{Nekrasov:2009rc, Mironov:2009uv})\footnote{For a recent application of the NS-limit to monopoles and vortices in the Higgs-phase, see \cite{Tong:2015kpa}.} such that $\epsilon_{+}=\epsilon_-=\epsilon_1/2$. Indeed, in this case, we have
\bea
\lim_{\epsilon_2\to 0} r_{\vec{\nu}}(m,\epsilon_1,\epsilon_2)=K\,m^2-\frac{K}{4}\epsilon_{1}^2\,,
\eea
which depends only on $\{\nu_1,\ldots,\nu_{N}\}$ through $K$ and hence can be absorbed in $Q_{f_{a}}$, making the partition function modular invariant without holomorphic anomaly at the expense of making $t_{f_{a}}$ transform as:
\bea
t_{f_{a}}\mapsto t_{f_{a}}-(m^2-\frac{\epsilon_{1}^2}{4})\,.
\eea
In our previous work \cite{Hohenegger:2015cba}, we gave a physical interpretation for the necessity of the NS limit when comparing BPS counting functions of M- and monopole-string excitations  (see also \cite{Haghighat:2015coa}).

\subsubsection{Transformation $\rho\mapsto \frac{a\rho+b}{c\rho+d}$}

Now let us  consider the transformation with respect to $\rho$ given by (\ref{ModTrans1}). To study it, we use the form of the topological string partition function of $X_{N}$ given by Eq.(\ref{GTPF1}),
%
\bea \label{GTPF1here}
\mathcal{Z}_{X_{N}}(\tau,m,t_{f_{1}},\cdots,t_{f_{N}},\epsilon_{1,2})=Z_{1}(m,t_{f_{1}},\cdots,t_{f_{N}},\epsilon_{1,2})\,\widetilde{\cal Z}^{(1)}_{N}(\tau,m,t_{f_{1}},\cdots,t_{f_{N}},\epsilon_1,\epsilon_2)\, . 
\eea
We recall from (\ref{ExpandInstanton}) that the  function $\widetilde{\mathcal{Z}}^{(1)}_{N}(\tau,m,t_{f_{1}},\cdots,t_{f_{N}},\epsilon_{1},\epsilon_{2})$ is given by 
\bea
\widetilde{\cal Z}^{(1)}_{N}
=\sum_{\alpha_{1}\cdots \alpha_{N}}Q_{\tau}^{|\alpha_{1}|+\cdots+|\alpha_{N}|}\,\prod_{a=1}^{N}\frac{\vartheta_{\alpha_{a}\alpha_{a}}(Q_{m})}{\vartheta_{\alpha_{a}\alpha_{a}}(\sqrt{\frac{t}{q}})}
\prod_{1\leq a<b\leq N}\frac{\vartheta_{\alpha_{a}\alpha_{b}}(Q_{ab}Q_{m}^{-1})\vartheta_{\alpha_{a}\alpha_{b}}(Q_{ab}Q_{m})}
{\vartheta_{\alpha_{a}\alpha_{b}}(Q_{ab}\sqrt{\frac{t}{q}})\vartheta_{\alpha_{a}\alpha_{b}}(Q_{ab}\sqrt{\frac{q}{t}})}\, \label{Z1here}
\eea
whose building blocks are the product of $\theta_{1}(\tau, z)$ functions:
\bea
\vartheta_{\mu\nu}(x)&=&\prod_{(i,j)\in \mu}\theta_1(\rho;x^{-1}t^{-\nu^{t}_{j}+i-\frac{1}{2}}q^{-\mu_{i}+j-\frac{1}{2}})\prod_{(i,j)\in \nu}\theta_1(\rho;x^{-1}t^{\mu^{t}_{j}-i+\frac{1}{2}}q^{\nu_{i}-j+\frac{1}{2}})\,. 
\eea
%
Since it is a sum over products of $\theta_{1}(\tau,z)$, as discussed in Section (\ref{SS1}), it too can be made modular invariant at the expense of introducing a holomorphic anomaly.  The function $Z_{1}$ in Eq.(\ref{GTPF1here}) has many properties similar to $W_{\emptyset\emptyset}$. In recent study \cite{Shabbir:2015oxa}, it was shown that $Z_1$ is modular invariant up to non-perturbative corrections in $\Omega$-deformation parameters 
in the refined topological string setup). Thus the complete partition function $\mathcal{Z}_{X_{N}}(\tau,m,t_{f_{1}},\cdots,t_{f_{N}},\epsilon_{1,2})$ is invariant under the modular transformation Eq.(\ref{ModTrans1}).

So far, we showed that the topological string partition function ${\cal Z}_{X_N}$ can be made fully modular invariant with respect to $\rho$ or $\tau$.  These two K\"ahler parameters are independent,  so it is expected that ${\cal Z}_{X_N}$ can be made simultaneously modular invariant with respect to both $\rho$ and $\tau$. We will not discuss technical details of the construction here except remarking that a closely parallel question was answered affirmatively positive in the context of topological string amplitudes of Type II string theory compactified on a two-parameter model of elliptically fibered Calabi-Yau threefolds  \cite{deWit:1996wq, deWit:1997ad}.

\section{Compact versus Non-Compact Free Energies}\label{Sect:RelCompNonComp}
We start by searching for relations between the BPS counting functions $F^{(\{k_i\})}(\tau,m,\epsilon_{1},\epsilon_2)$ of the non-compact theory and $G^{(\{k_i\})}(\tau,m,\epsilon_1,\epsilon_2)$ of gauge theory 2. We first consider the special class of configurations $\{k_i\}=\{1,\ldots,1\}$ and conjecture the relation for the generic case based on an emergent pattern. We also comment on implications of this pattern on the little string theories. 
\subsection{Examples of Compact Free Energies $G^{(k_1,\ldots,k_N)}$}
The simplest configuration in the compact case corresponds to a single M2-brane starting and ending on the same M5-brane.  In our notation, this corresponds to $N=1$ and $\{k_i\}=(1)$. The BPS bound states of this configuration are counted by
\bea
G^{(1)}(\tau,m,\epsilon_{1},\epsilon_2)=\frac{\theta_{1}(\tau;m+\epsilon_{-})\theta_{1}(\tau;m-\epsilon_{-})}{\theta_{1}(\tau;\epsilon_{1})\theta_{1}(\tau;\epsilon_{2})}\,.\label{G1com}
\eea
On the other hand, the simplest configuration in the non-compact case corresponds to a single M2-brane stretched between two M5-branes, for which the corresponding BPS-counting function is given by
\bea
F^{(1)}(\tau,m,\epsilon_1,\epsilon_2)=\frac{\theta_{1}(\tau;m+\epsilon_{+})\theta_{1}(\tau;m-\epsilon_{+})}{\theta_{1}(\tau;\epsilon_{1})\theta_{1}(\tau;\epsilon_{2})}\,.\label{F1noncom}
\eea
Comparing (\ref{G1com}) with (\ref{F1noncom}), we notice the following relation
\bea\label{NSrelation}
G^{(1)}(\tau,m,\epsilon_1,\epsilon_2)=-F^{(1)}(\tau,m,\epsilon_1,-\epsilon_2)=-F^{(1)}(\tau,m,-\epsilon_1,\epsilon_2)\,.
\eea
Most importantly, while both $F^{(1)}$ and $G^{(1)}$ have a first order pole for $\epsilon_2=0$, we find in the NS-limit the relation
\begin{align}
\lim_{\epsilon_2\to 0}\,\epsilon_2\,G^{(1)}(\tau,m,\epsilon_1,\epsilon_2)=\lim_{\epsilon_2\to 0}\,\epsilon_2\,F^{(1)}(\tau,m,\epsilon_1,\epsilon_2)\,,\label{RelCompNonComp1}
\end{align}
which we will use later on.

The next, more complicated configuration is $G^{(1,1)}$, which corresponds to two M5-branes with two M2-branes stretched between them. Their BPS-counting function is given by
\begin{align}
G^{(1,1)}&=\Big(\frac{\theta_{1}(\tau;m+\epsilon_{-})\theta_{1}(\tau;m-\epsilon_{-})}{\theta_{1}(\tau;\epsilon_{1})\theta_{1}(\tau;\epsilon_{2})}\Big)^2-\Big(\frac{\theta_{1}(\tau;m+\epsilon_{+})\theta_{1}(\tau;m-\epsilon_{+})}{\theta_{1}(\tau;\epsilon_{1})\theta_{1}(\tau;\epsilon_{2})}\Big)^2\,.\label{DefConf11comp}
\end{align}
Further configurations can be worked out in the same manner. However, their free energies are generically very complicated and we will not display them here in full generality. 

Following the reasoning in our previous paper~\cite{Hohenegger:2015cba} for the non-compact free energies $F^{(\{k_i\})}$, we will consider the NS-limit together with a series expansion in the remaining deformation parameter $\epsilon_1$
\begin{align}
&\lim_{\epsilon_2\to 0}\frac{G^{(\{k_i\})}(\tau,m,\epsilon_1,\epsilon_2)}{G^{(1)}(\tau,m,\epsilon_1,\epsilon_2)}=\sum_{n=0}^\infty \epsilon_1^{2n}\,g^{n,(\{k_i\})}(\tau,m)\,,\label{SeriesGNS}\\
&\lim_{\epsilon_2\to 0}\frac{F^{(\{k_i\})}(\tau,m,\epsilon_1,\epsilon_2)}{F^{(1)}(\tau,m,\epsilon_1,\epsilon_2)}=\sum_{n=0}^\infty \epsilon_1^{2n}\,f^{n,(\{k_i\})}(\tau,m)\,.
\end{align}
Dividing by $F^{(1)}$ and $G^{(1)}$, respectively, removes the $\epsilon_2^{-1}$ pole and yields a finite NS-limit. Furthermore, the coefficient functions $g^{n,(\{k_i\})}$ and $f^{n,(\{k_i\})}$ are quasi-modular Jacobi forms of weight $2n$ and index $K=\sum_a k_a$, \emph{i.e.} they can be written in the following form
\begin{align}
&g^{n,(\{k_i\})}(\tau,m)=\sum_{a=0}^K s^{(n,\{k_i\})}_{2a+2n}(\tau)\,\left(\varphi_{0,1}(\tau,m)\right)^{K-a}\,\left(\varphi_{-2,1}(\tau,m)\right)^{a}\,,\\
&f^{n,(\{k_i\})}(\tau,m)=\sum_{a=0}^K t^{(n,\{k_i\})}_{2a+2n}(\tau)\,\left(\varphi_{0,1}(\tau,m)\right)^{K-a}\,\left(\varphi_{-2,1}(\tau,m)\right)^{a}\,.
\end{align}
Here, $s^{(n,\{k_i\})}_{m}$ and $t^{(n,\{k_i\})}_{m}$ are quasi-modular forms of weight $m$, which can be written as polynomials in Eisenstein series (including $E_2(\tau)$). The explicit expressions for a few $f^{n,(\{k_i\})}$ and $g^{n,(\{k_i\})}$ for simple configurations $(\{k_i\})$ are given in appendix~\ref{App:SerExpansion}.

Eq. (\ref{RelCompNonComp1}) shows that the free energies of the simplest compact and non-compact configurations of M5-branes agree in the NS limit. In the following, we address the question whether there are further relations between $G^{(\{k_i\})}$ and $F^{(\{k_i\})}$ for more complicated configurations $\{k_i\}$.
\subsection{Configurations $(1,\ldots,1)$}
In \cite{Hohenegger:2015cba}, we have seen that the free energies for configurations $(\{k_i\})=(\underbrace{1,\ldots,1}_{N-1 \text{ times}})$, \emph{i.e.} for $N$ parallel M5-branes with a single M2-brane between each of them in the non-compact case are proportional to $F^{(1)}$. Specifically, they can be written in the form
\begin{align}
F^{(1,\ldots,1)}(\tau,m,\epsilon_1,\epsilon_2)=F^{(1)}(\tau,m,\epsilon_1,\epsilon_2)\,W(\tau,m,\epsilon_1,\epsilon_2)^{N-2}\,,
\end{align}
with
\begin{align}
W(\tau,m,\epsilon_1,\epsilon_2)=\frac{\theta_1(\tau;m+\epsilon_-)\theta_1(\tau;m-\epsilon_-)-\theta_1(\tau;m+\epsilon_+)\theta_1(\tau;m-\epsilon_+)}{\theta_1(\tau;\epsilon_1)\theta_1(\tau;\epsilon_2)}\,.
\end{align}
We therefore expect that the counting function for configurations with $N$ M5-branes on a circle with a single M2-brane between each of them should also simplify in the NS-limit. The first non-trivial such configuration is $G^{(1,1)}$ introduced in (\ref{DefConf11comp}). It can be written in the following manner
\begin{align}
G^{(1,1)}(\tau,m,\epsilon_{1},\epsilon_2)&=\Big(G^{(1)}(\tau,m,\epsilon_{1},\epsilon_2)\Big)^2-\Big(G^{(1,0)}(\tau,m,\epsilon_{1},\epsilon_2)\Big)^2\nonumber\\
&=W(\tau,m,\epsilon_{1},\epsilon_2)\,\Big[G^{(1)}(\tau,m,\epsilon_{1},\epsilon_2)+G^{(1,0)}(\tau,m,\epsilon_{1},\epsilon_2)\Big]\nonumber\\
&=W(\tau,m,\epsilon_{1},\epsilon_2)\,\Big[G^{(1)}(\tau,m,\epsilon_{1},\epsilon_2)+F^{(1)}(\tau,m,\epsilon_{1},\epsilon_2)\Big]\,,\label{Example11}
\end{align}
where in the last line we have used (\ref{DefZline}). Using furthermore (\ref{RelCompNonComp1}), this relation simplifies in the NS-limit
\begin{align}
\lim_{\epsilon_2\to 0}\frac{G^{(1,1)}(\tau,m,\epsilon_{1},\epsilon_2)}{G^{(1)}(\tau,m,\epsilon_{1},\epsilon_2)}=2\,W(\tau,m,\epsilon_1,\epsilon_2=0)\,.
\end{align}

\noindent
The next more complicated configuration is $(1,1,1)$ for which we find
{\allowdisplaybreaks\begin{align}
G^{(1,1,1)}(\tau,m,\epsilon_{1},\epsilon_2)&=\Big(G^{(1)}(\tau,m,\epsilon_{1},\epsilon_2)\Big)^3-3G^{(1)}(\tau,m,\epsilon_{1},\epsilon_2)\,\Big(G^{(1,0)}(\tau,m,\epsilon_{1},\epsilon_2)\Big)^2\nonumber\\
&\hspace{2cm}+2\Big(G^{(1,0)}(\tau,m,\epsilon_{1},\epsilon_2)\Big)^3\nonumber\\
&=W(\tau,m,\epsilon_{1},\epsilon_2)^2\,\Big[G^{(1)}(\tau,m,\epsilon_{1},\epsilon_2)+ 2\,G^{(1,0)}(\tau,m,\epsilon_{1},\epsilon_2)\Big]\nonumber\\
&=W(\tau,m,\epsilon_{1},\epsilon_2)^2\,\Big[G^{(1)}(\tau,m,\epsilon_{1},\epsilon_2)+ 2\,F^{(1)}(\tau,m,\epsilon_{1},\epsilon_2)\Big]\,.\label{Example111}
\end{align}}
\noindent 
Generalizing the two examples (\ref{Example11}) and (\ref{Example111}) we conjecture the general pattern,
\begin{align}\label{simpleconfiguration}
G^{(1,\cdots,1)}(\tau,m,\epsilon_{1},\epsilon_2)&=W(\tau,m,\epsilon_{1},\epsilon_2)^{N-1}\Big[G^{(1)}(\tau,m,\epsilon_{1},\epsilon_2)+(N-1)G^{(1,0)}(\tau,m,\epsilon_{1},\epsilon_2)\Big]\nonumber\\
&=W(\tau,m,\epsilon_{1},\epsilon_2)^{N-1}\Big[G^{(1)}(\tau,m,\epsilon_{1},\epsilon_2)+(N-1)F^{(1)}(\tau,m,\epsilon_{1},\epsilon_2)\Big]\,.
\end{align}
Thus, the counting of a circular M2-brane over $N$ intervals can be generated from the counting of a circular M2-brane over $(N-1)$ intervals via the two-term recursion relation: 
\vskip0.3cm
\begin{tcolorbox}
${}$\\[-30pt]
\begin{align}
G^{\overbrace{\text{{\scriptsize{$(1, \cdots, 1)$}}}}^{N\ \ \rm times}} (\tau, m, \epsilon_1, \epsilon_2)
&= W (\tau, m, \epsilon_1, \epsilon_2) G^{\overbrace{\text{{\scriptsize{$(1, \cdots, 1)$}}}}^{(N -1) \ \rm times}} (\tau, m, \epsilon_1, \epsilon_2) \qquad \qquad \qquad (N > 1)
\nonumber \\\label{recursionrelation}
&+ W(\tau, m, \epsilon_1, \epsilon_2)^{(N-1)}  F^{(1)}(\tau, m, \epsilon_1, \epsilon_2) \ .\\\nonumber
\end{align}
${}$\\[-60pt]
\end{tcolorbox}
\noindent 
We checked (\ref{recursionrelation}) explicitly up to $N=5$. A different way to express the recursion relation in Eq.(\ref{recursionrelation}) is the following:
\bea\nonumber
G^{\overbrace{\text{{\scriptsize{$(1, \cdots, 1)$}}}}^{N\ \ \rm times}} (\tau, m, \epsilon_1, \epsilon_2)=W (\tau, m, \epsilon_1, \epsilon_2) \Big(G^{\overbrace{\text{{\scriptsize{$(1, \cdots, 1)$}}}}^{(N -1) \  \rm times}} (\tau, m, \epsilon_1, \epsilon_2) + F^{\overbrace{\text{{\scriptsize{$(1, \cdots, 1)$}}}}^{(N -1) \  \rm times}} (\tau, m, \epsilon_1, \epsilon_2) \Big)\,.
\eea

\vskip0.3cm
\subsection{General Configurations}
While the counting functions $G^{(1,\ldots,1)}$ reduce to a universal structure in the NS-limit, more general configurations show more involved relations to the non-compact $F^{ (\{k_i\} )}$'s.  To study these configurations, we may work perturbatively in $\epsilon_1$. Indeed, in the NS-limit, we have worked out several $G^{(\{k_i\})}$ and $F^{(\{k_i\})}$ in appendix~\ref{App:RelsCompNonComp} to various orders in $\epsilon_1$. Built upon these examples, we conjecture a general pattern for these relations.

For a general BPS configuration $G^{(\{k_i\})}$, labelled by the sequence of positive integers $(\{k_i\})=(k_1,\ldots,k_\ell)$ with $\sum_{i=1}^\ell k_i = K$ and $k_{a=1,\cdots,\ell}\neq 0$, we find that \\[4pt]

\begin{tcolorbox}
${}$\\[-30pt]
\begin{align}
\label{generalconfig}
G^{(\{k_i\})}(\tau,m,\epsilon_{1},\epsilon_2)=d_{(\{k_i\})}\sum_{\sum m_i=K} a_{(\{m_i\})}\,F^{(\{m_i\})}(\tau,m,\epsilon_{1},\epsilon_2)\,,
\end{align}
\end{tcolorbox}
${}$\\
\noindent
{is compatible with all cases worked out in appendix~\ref{App:RelsCompNonComp}. Here, the summation is over all sequences of positive integers $(\{m_i\})=(m_1,\cdots,m_p)$ such that $\sum_{a=1}^pm_a=K$ and  $a_{(\{m_i\})}$ and $d_{(\{k_i\})}$ are integer-valued coefficients that depend on the combinatorics of the $(\{k_i\})$ and $(\{m_i\})$, respectively.  

Specifically, the prefactor $d_{(\{k_i\})}$ is non-trivial (\emph{i.e.} it differs from $1$) if the corresponding $(\{k_i\})$ can be written as an iteration of a smaller (elementary) building block $\{k_j\}_m=(k_1,\ldots,k_m)$ with $m<\ell$ and $n=\frac{\ell}{m}\in\mathbb{N}$
\begin{align}
d_{(\{k_i\})}=\left\{\begin{array}{lcl} n = {\ell \over m} & \text{if} & (\{k_i\}) = ( \underbrace{\{k_j\}_m ,\ldots,\{k_j\}_m}_{n\text{ times}} ) \\[-18pt] \\ 1 & \  \  &\text{else} \end{array}\right.\label{Prefact}
\end{align}
For example, the configuration $(2,1,2,1)$ is the double repetition of the elementary block $(2,1)$ and $(1,1,1)$ is the threefold repetition of the elementary block $(1)$, while $(2,2,1)$ cannot be written as the iteration of a more elementary block:
\begin{align}
&d_{(2,1,2,1)}=2\,,&&d_{(1,1,1)}=3\,,&&d_{(2,2,1)}=1\,.
\end{align}  
The relative coefficients $a_{(\{m_i\})}$ single out specific configurations $(\{m_i\})$
\begin{align}\label{Relfact}
a_{(\{m_i\})}=\left\{\begin{array}{lclc} 1 & \text{if} & k_i=\sum_{r=0}^\infty m_{i+r\ell}  \quad (i = 1, \cdots, \ell) \\ 0 & \text{else} \end{array}\right. \, . 
\end{align}
We can describe this prescription in a more intuitive way: the idea is to construct the compact sequence $(\{k_i\})=(k_1,\ldots,k_\ell)$ by 'tape-wrapping' the non-compact sequence $(\{m_i\})=(m_1,\ldots,m_p)$ multiple times around a circle of circumference $\ell$: 
\begin{center}
\begin{tikzpicture}
\draw[ultra thick] (0,0) circle (1.5cm);
\node at (0,1.2) {$k_1$};
\node at (0.8,0.8) {$k_2$};
\node at (1.2,0) {$k_3$};
\node[rotate around={45:(0,0)}] at (0.8,-0.8) {$\cdots$};
\node at (-0.8,0.8) {$k_\ell$};
\node at (-0.9,0) {$k_{\ell-1}$};
\node at (0,1.7) {$m_1$};
\node at (1.3,1.3) {$m_2$};
\node at (1.9,0) {$m_3$};
\node[rotate around={45:(0,0)}] at (1.3,-1.3) {$\cdots$};
\node at (-2,0) {$m_{\ell-1}$};
\node at (-1.4,1.4) {$m_{\ell}$};
\node at (0,2.2) {$m_{\ell+1}$};
\node at (1.8,1.8) {$m_{\ell+2}$};
\node at (2.9,0) {$m_{\ell+3}$};
\node[rotate around={45:(0,0)}] at (1.8,-1.8) {$\cdots$};
\node at (-3.1,0) {$m_{2\ell-1}$};
\node at (-1.9,1.9) {$m_{2\ell}$};
\node at (0,2.9) {$\vdots$};
\node[rotate around={45:(0,0)}] at (2.3,2.3) {$\cdots$};
\node at (3.7,0) {$\cdots$};
\node[rotate around={-45:(0,0)}] at (2.3,-2.3) {$\cdots$};
\node at (-4,0) {$\cdots$};
\node[rotate around={-45:(0,0)}] at (-2.4,2.4) {$\cdots$};
\node at (0,3.3) {$m_{s\ell+1}$};
\node at (2.8,2.8) {$m_{s\ell+2}$};
\node at (4.5,0) {$m_{s\ell+3}$};
\node[rotate around={45:(0,0)}] at (2.8,-2.8) {$\cdots$};
\node at (-5.2,0) {$m_{(s+1)\ell-1}$};
\node at (-2.9,2.9) {$m_{(s+1)\ell}$};
\node at (0,3.8) {$m_{p-2}$};
\node at (3.2,3.2) {$m_{p-1}$};
\node at (5.5,0) {$m_{p}$};
%
%
\end{tikzpicture}
\end{center}
The coefficient $a_{(\{m_i\})}$ is non-zero only, if the overlapping BPS excitations of the noncompact $(\{m_i\})$ add up to the $k_i$ of the compact BPS excitations }
\begin{align}
&k_1= m_1+m_{\ell+1}+\ldots\,,&&k_2=m_2+m_{\ell+2}+\ldots\,,&&\text{etc.}\,,&&k_\ell=m_\ell+m_{2\ell}+\ldots\,.
\end{align}
In the figure above, this corresponds to summing up all multiplicities along the radial directions.

We note that this 'wrapping' prescription also reproduces the correct relation between $G^{(\{k_i\})}$ and $F^{(\{m_i\})}$ if one of the $k_i$ vanishes. Due to the cyclic symmetry of the partition $k_i$, we can without loss of generality choose $k_\ell=0$. In this case, the conditions we obtain from the wrapping procedure are
\begin{align}
&k_1= m_1+m_{\ell+1}+\ldots\,,&&k_2=m_2+m_{\ell+2}+\ldots\,,&&\text{etc.}\,,&&0=m_\ell+m_{2\ell}+\ldots\,.
\end{align}
This in particular indicates that $m_\ell=0$, which means that the non-compact configuration $\{m_i\}$ has only $\ell-1$ entries (\emph{i.e.} it does not fully wrap around the compact configuration). Therefore, the only configuration contributing is
\begin{align}
&m_i=k_i\,,&&\forall i=1,\ldots,\ell-1\,.
\end{align}
The coefficient in this case, however, is always $1$:
\begin{align}
&G^{(\{k_1,\ldots,k_{\ell-1},0\})}=F^{(\{k_1,\ldots,k_{\ell-1}\})}\,.
\end{align}
Finally, let us illustrate the procedure with an example: consider $G^{(3,1)}$ (with $\ell=2$). According to (\ref{Prefact}), we have $d_{(3,1)}=1$. Furthermore, there are eight compact $F^{(\{m_i\})}$ with $\sum_i m_i=4$
\begin{align}
&F^{(4)}\,,&&F^{(3,1)}=F^{(1,3)}\,,&&F^{(2,2)}\,,&&F^{(2,1,1)}=F^{(1,1,2)}\,,&&F^{(1,2,1)}\,,&&F^{(1,1,1,1)}\,.
\end{align}
For each of these $\{m_i\}$, we can compute the sum $\sum_{r}m_{i+2r}$, which we can tabulate as follows
\begin{center}
\begin{tabular}{|c|c|c|}\hline
&&\\[-14pt]
$\{m_i\}$ & $\left\{\sum_{r}m_{i+2r}\right\}$ & $a_{(\{m_i\})}$\\[2pt]\hline\hline
&&\\[-14pt]
$(4)$ & $(4)$ & $0$\\[2pt]\hline
&&\\[-14pt]
$(3,1)$ & $(3,1)$ & $1$\\[2pt]\hline
&&\\[-14pt]
$(2,2)$ & $(2,2)$ & $0$\\[2pt]\hline
&&\\[-14pt]
$(2,1,1)$ & $(3,1)$ & $1$\\[2pt]\hline
\end{tabular}\hspace{1cm}
\begin{tabular}{|c|c|c|}\hline
&&\\[-14pt]
$\{m_i\}$ & $\left\{\sum_{r}m_{i+2r}\right\}$ & $a_{(\{m_i\})}$\\[2pt]\hline\hline
&&\\[-14pt]
$(1,3)$ & $(1,3)$ & $1$\\[2pt]\hline
&&\\[-14pt]
$(1,1,2)$ & $(3,1)$ & $1$\\[2pt]\hline
&&\\[-14pt]
$(1,2,1)$ & $(2,2)$ & $0$\\[2pt]\hline
&&\\[-14pt]
$(1,1,1,1)$ & $(2,2)$ & $0$\\[2pt]\hline
\end{tabular}
\end{center}
These are indeed the coefficients we find in the genus expansion in (\ref{RelsK4}).

As another example consider the configuration $(2,1,2,1)$. From (\ref{Prefact}) and (\ref{Relfact}) it follows that $m=2$ and
\bea
G^{(2,1,2,1)}&=&2\Big(F^{(2,1,2,1)} +F^{(1,2,1,2)}+ F^{(1,1,2,1,1)}\Big)\,,\\\nonumber
&=&2\Big(2F^{(2,1,2,1)}+F^{(1,1,2,1,1)}\Big),
\eea
where the second equation follows from the fact that $F^{(1,2,1,2)}=F^{(2,1,2,1)}$.

\section{Monopole versus Instanton Free Energies}\label{Sect:MonopolesvsInst}
In this section, we discover remarkable relations between the counting function of gauge theory 1 and the counting function of gauge theory 2.
\subsection{Connection between Monopole and Instanton Free Energies}
The moduli $(t_{f_1},\ldots,t_{f_{N-1}})$ transform in a non-trivial fashion with respect to (\ref{ModTrans1}). Therefore, the coefficients $\Sigma_{N,k}^{(k_1,\ldots,k_{N-1})}(\rho,m,\epsilon_1,\epsilon_2)$ (see (\ref{SigmaDef})) generically do not transform nicely under (\ref{ModTrans1}). In the NS limit $\epsilon_2\to 0$, the function  $\Sigma_{N,k}(\rho,m,t_{f_{1}},\cdots,t_{f_{N-1}},\epsilon_{1,2})$ in (\ref{SigmaDef}) transforms with index $k$ for each $t_{f_{a}}$ under (\ref{ModTrans1}) and hence can be re-expressed as an expansion in terms of a basis~\footnote{For a definition of the $\vartheta_{s,m}(\rho,t_{f_i})$, we refer readers to our previous paper\cite{Hohenegger:2015cba}.} of theta-functions of index $k$. Thus, 
\begin{align}
\lim_{\epsilon_2\to 0}\epsilon_2\Sigma_{N,k}(\rho,& m,t_{f_1},\ldots,t_{f_{N-1}},\epsilon_{1},\epsilon_{2})\nonumber\\
&=\sum_{m_1=0}^{2k-1}\ldots \sum_{m_{N-1}=0}^{2k-1}\vartheta_{k,m_1}(\rho,t_{f_1})\ldots\vartheta_{k,m_{N-1}}(\rho,t_{f_{N-1}})\,h_{m_1,\ldots,m_{N-1}}(\rho,m,\epsilon_1) \, , 
\end{align}
and the coefficients $h_{m_1,\ldots,m_{N-1}}$ will transform as vector-valued modular forms under the $SL(2,\mathbb{Z})$ transformation generated by
\begin{align}
(\rho,m,\epsilon_1)&\mapsto \Big(\frac{a\rho+b}{c\rho+d},\frac{m}{c\rho+d},\frac{\epsilon_1}{c\rho+d}\Big)\,.
\end{align}
However, they may transform covariantly under certain congruence subgroups. Therefore, the coefficients $h_{m_1,\ldots,m_{N-1}}$ have the properties that allow them to be compared with the free energies of certain monopole string configurations. To check this, we extract the simplest coefficient $h_{0,\ldots,0}(\rho,m,\epsilon_1)=\lim_{\epsilon_2\to 0}\epsilon_2 \sigma_{N,k}(\rho,m,\epsilon_1,\epsilon_2)$ through
\begin{align}
\sigma_{N,k}(\rho,m,\epsilon_1,\epsilon_2):=\Sigma_{N,k}^{(0,\ldots,0)}(\rho,m,\epsilon_1,\epsilon_2)=
\oint_0 \frac{dQ_{f_1}}{Q_{f_1}}\ldots \oint_0 \frac{dQ_{f_{N-1}}}{Q_{f_{N-1}}}
\,\Sigma_{N,k}(\rho,m,t_{f_1},\ldots, t_{f_{N-1}},\epsilon_1,\epsilon_2)\, ,  \label{Defsigma}
\end{align}
where the contour integrals~\footnote{ The $i$th contour is defined as a small circle around the point $Q_{f_i}=0$, as was previously prescribed in non-compact situation in \cite{Hohenegger:2015cba} (see also similar considerations in \cite{Kim:2011mv}).} are just to extract the constant term of $\Sigma_{N, k}(\rho, m, t_{f_1}, \cdots, t_{f_{N-1}}, \epsilon_{1,2})$ in an expansion of the fugacities $Q_{f_1}, \cdots, Q_{f_{N-1}}$, as defined in Eq.(\ref{SigmaDef}). 
Remarkably, in the NS limit, we find evidence that the quotients $(\sigma_{N,k}/\sigma_{1,1})$ are related to the free energies of specific configurations $(\{k_i\})$ of the monopole strings with $k_1=k_2=\ldots=k_N=k$. Indeed, based on the examples discussed below, we conjecture \\
\begin{tcolorbox}
${}$\\[-30pt]
\begin{align}
\lim_{\epsilon_2\to0}\,\frac{\sigma_{N,k}(t,m,\epsilon_1,\epsilon_2)}{\sigma_{1,1}(t,m,\epsilon_1,\epsilon_2)}=\lim_{\epsilon_2\to 0}\,\frac{G^{\overbrace{\text{{\scriptsize{$(k, \ldots, k)$}}}}^{N \ \rm times}}(t,m,\epsilon_1,\epsilon_2)}{G^{(1)}(t,m,\epsilon_1,\epsilon_2)}\,.\label{ConjectureCon}
\end{align}
\end{tcolorbox}
\vskip0.3cm
\noindent 
Interpreting this conjecture from the point of view of IIa and IIb little strings (see section~\ref{Sect:LittleStrings}) we notice that $\sigma_{N,k}$ and $G^{(k,\ldots,k)}$ are the free energies of configurations of little strings with momentum and winding number $k$ in type IIa and IIb respectively.  Thus, we believe our conjecture is in line with the T-duality property between type IIa and IIb little strings. Indeed, under T-duality, the momentum quantum number $k$, weighed with the fugacity $Q_\tau^k$, is mapped to the winding quantum number $k$, weighed with the fugacity $Q_\rho^k$. 

\subsection{Checks and Series Expansions}
In this subsection, we provide support for the conjecture (\ref{ConjectureCon}): we give an analytic proof for the case $k=1$ (and $N$ generic) and provide additional checks for $k>1$ by comparing the power series expansion of the left- and right-hand sides of (\ref{ConjectureCon}).
\subsubsection{The case $C_{N,1}$}\label{Sect:PropertyPart}
To simplify the notation, we introduce the short-hand for the individual building blocks in the partition function (\ref{Z1part}):
\bea
E(\rho,t,m,\epsilon):=\frac{\theta_1(\rho;t+m)\theta_1(\rho;t-m)}{\theta_1(\rho;t+\epsilon)\theta_1(\rho;t-\epsilon)}\,.
\eea
Using these building blocks, we can write
\begin{align}
\lim_{\epsilon_2\to 0}\frac{C_{N,1}(\rho,m,t_{f_{1}},\cdots,t_{f_{N-1}},\epsilon_1,\epsilon_2)}{C_{1,1}(\rho,m,\epsilon_1,\epsilon_2)}&=\sum_{k=1}^{N}\prod_{a=1}^{k-1}E(\rho,t_{ak}-\epsilon_{+},m,\epsilon_{+})\,\prod_{b=k+1}^{N}E(\rho,t_{kb}+\epsilon_{+},m,\epsilon_{+})\nonumber\\
&=\sum_{k=1}^{N}\prod_{a=1}^{k-1}E(\rho,\widehat{t}_{ak},m,\epsilon_{+})\,\prod_{b=k+1}^{N}E(\rho,\widehat{t}_{kb},m,\epsilon_{+})\, , 
\end{align}
where we introduced the shorthand
\begin{align}
&\widehat{t}_{ak}=t_{ak}-\epsilon_{+}\,,&&\text{for} &&a=1,\cdots,k-1\,,\\\nonumber
&\widehat{t}_{kb}=t_{kb}+\epsilon_{+}\,,&&\text{for} && b=k+1,\cdots,N\,.
\end{align}
Following (\ref{Defsigma}) and (\ref{ConjectureCon}), we are interested in the terms that are independent in $Q_{f_{a}}$, which are extracted by contour integration:
\bea
\oint \frac{dQ_{f_{1}}\cdots dQ_{f_{N-1}}}{Q_{f_{1}}\cdots Q_{f_{N-1}}}\,\left(\lim_{\epsilon_2\to 0 }\frac{C_{N,1}(\rho,m,t_{f_{1}},\cdots,t_{f_{N-1}},\epsilon_1,\epsilon_2)}{C_{1,1}(\rho,m,\epsilon_1,\epsilon_2)}\right)\,.
\eea
Using the definition $Q_{f_a}=e^{-2 \pi t_{f_a}}$, we can perform the following change of variables
\begin{align}
&\oint \frac{dQ_{f_{1}}\cdots dQ_{f_{N-1}}}{Q_{f_{1}}\cdots Q_{f_{N-1}}}=(-1)^{N-1}\int dt_{12}dt_{23}\cdots dt_{N-1\,N}=(-1)^{N-1}\int \prod_{a=1}^{k-1}dt_{ak}\,\,\prod_{b=k+1}^{N}dt_{kb}\,,&&\forall k\,.
\end{align}
Shifting the individual $t_{ak}$ and $t_{kb}$, we then find
\begin{align}
\oint \frac{dQ_{f_{1}}\cdots dQ_{f_{N-1}}}{Q_{f_{1}}\cdots Q_{f_{N-1}}}\,&\left(\lim_{\epsilon_2\to 0 }\frac{C_{N,1}(\rho,m,t_{f_{1}},\cdots,t_{f_{N-1}},\epsilon_1,\epsilon_2)}{C_{1,1}(\rho,m,\epsilon_1,\epsilon_2)}\right)\nonumber\\
&=(-1)^{N-1}N\,\Big(\int dt\, E(\rho,t,m,\epsilon_{1})\Big)^{N-1}\nonumber\\
&=N\,(\lim_{\epsilon_{2}\mapsto 0}W(\rho,m,\epsilon_{1},\epsilon_2))^{N-1}\,,\label{RelResidue}
\end{align}
where we have used the relation ($x=e^{2\pi i\,t}$),
\bea
\oint\,\frac{dx}{2\pi ix}\,\frac{\theta_{1}(\rho;t+m)\theta_{1}(\rho;t-m)}{\theta_{1}(\rho;t+\epsilon_1)\theta_{1}(\rho;t-\epsilon_1)}&=&\lim_{\epsilon_{2}\mapsto 0}W(\rho,m,\epsilon_1,\epsilon_2)\\\nonumber
&=&\frac{\theta_{1}(\rho;m-\frac{\epsilon_1}{2})\theta'_{1}(\rho;m+\frac{\epsilon_1}{2})-\theta_{1}(\rho;m+\frac{\epsilon_1}{2})\theta'_{1}(\rho;m-\frac{\epsilon_1}{2})}{\theta_{1}(\rho;\epsilon_1)\theta'_{1}(\rho;0)}\,.
\eea
\noindent
We have additionally checked (\ref{RelResidue}) up to $N=4$ through an explicit computation of $C_{N,1}$.

\subsubsection{Case $C_{N,k>1}$}
For $k>1$, the quantities $\Sigma_{N,k}(\rho,m,t_{f_1},\ldots, t_{f_{N-1}},\epsilon_1,\epsilon_2)$ become complicated quotients of $\theta_1$-functions and we therefore only study their series expansions. Concretely, to compare with (\ref{SeriesGNS}), we introduce
\begin{align}
&\lim_{\epsilon_2\to0}\,\frac{\sigma_{N,k}(\rho,m,\epsilon_1,\epsilon_2)}{\sigma_{1,1}(\rho,m,\epsilon_1,\epsilon_2)}=\sum_{n=0}^\infty \epsilon_1^{2n}\,\sigma^{n}_{N,k}(\rho,m)\,.\label{SeriesSigmaNS}
\end{align}
Starting with $(N,k)=(2,2)$, we have for the cases $n=1,2$
{\allowdisplaybreaks\begin{align}
\sigma^{0}_{2,2}(\rho,m)&=2 \nonumber \\
&+Q_\rho \left(-4 Q_m ^3-\frac{4}{Q_m ^3}+38Q_m ^2+\frac{38}{Q_m ^2}-124 Q_m -\frac{124}{Q_m }+180\right)\nonumber\\
&+Q_\rho^2 \left(38Q_m ^4+\frac{38}{Q_m ^4}-448 Q_m ^3-\frac{448}{Q_m ^3}+2012 Q_m ^2+\frac{2012}{Q_m^2}-4640 Q_m -\frac{4640}{Q_m }+6076\right)\nonumber\\
&+Q_\rho^3 \bigg(2 Q_m ^6+\frac{2}{Q_m ^6}-124 Q_m ^5-\frac{124}{Q_m ^5}+2012
   Q_m ^4+\frac{2012}{Q_m ^4}-12892 Q_m ^3-\frac{12892}{Q_m ^3}\nonumber\\
&\hspace{1.2cm}+43350
   Q_m ^2+\frac{43350}{Q_m ^2}-86568 Q_m -\frac{86568}{Q_m }+108440\bigg) \nonumber \\
&+\mathcal{O}(Q_\rho^4)\,,\\
\sigma^{1}_{2,2}(\rho,m)&=Q_\rho \left(2 Q_m ^3+\frac{2}{Q_m ^3}-32Q_m ^2-\frac{32}{Q_m ^2}+158 Q_m +\frac{158}{Q_m }-264\right)\nonumber\\
&+Q_\rho^2 \left(-32 Q_m ^4-\frac{32}{Q_m ^4}+800Q_m ^3+\frac{800}{Q_m ^3}-4824 Q_m ^2-\frac{4824}{Q_m ^2}+12944Q_m +\frac{12944}{Q_m }-17792\right)\nonumber\\
&+Q_\rho^3 \bigg(158 Q_m ^5+\frac{158}{Q_m ^5}-4824 Q_m ^4-\frac{4824}{Q_m ^4}+42366Q_m ^3+\frac{42366}{Q_m ^3}-169920 Q_m ^2\nonumber\\
&\hspace{1.2cm}-\frac{169920}{Q_m ^2}+372708Q_m +\frac{372708}{Q_m }-481008\bigg)\nonumber \\
&+\mathcal{O}(Q_\rho^4)\,,
\end{align}}
which indeed agree with the expansion of (\ref{Expandg22}). 

For $(N,k)=(2,3)$, we find
\begin{align}
\sigma^0_{2,3}(\rho,m)&=2 \nonumber \\
&+Q_\rho \left(2
   Q_m ^4+\frac{2}{Q_m ^4}-60 Q_m ^3-\frac{60}{Q_m ^3}+360 Q_m ^2+\frac{360}{Q_m ^2}-944
   Q_m -\frac{944}{Q_m }+1284\right)\nonumber\\
&+Q_\rho^2 \bigg(2 Q_m ^6+\frac{2}{Q_m ^6}-200 Q_m ^5-\frac{200}{Q_m ^5}+3010
 Q_m ^4+\frac{3010}{Q_m ^4}-18396 Q_m ^3-\frac{18396}{Q_m ^3}\nonumber\\
 &\hspace{1.2cm}+60284
   Q_m ^2+\frac{60284}{Q_m ^2}-118840 Q_m -\frac{118840}{Q_m }+148280\bigg)\nonumber \\
& +\mathcal{O}(Q_\rho^3)\,,\\
\sigma^1_{2,3}(\rho,m)&=Q_\rho\,\frac{3 \left(3 Q_m ^6-28 Q_m ^5+103 Q_m ^4-158 Q_m ^3+103 Q_m ^2-28 Q_m +3\right)}{2 Q_m ^3} \nonumber \\
&+\mathcal{O}(Q_\rho^2)\,,
\end{align}
which indeed agrees with the corresponding expansions of $g^{0,(3,3)}$ and $g^{1,(3,3)}$ respectively. 

Finally, for $(N,k)=(3,2)$, we have
\begin{align}
\sigma^0_{3,2}(\rho,m)&=3 \nonumber \\
&+Q_\rho\left(3 Q_m ^4+\frac{3}{Q_m ^4}-36
   Q_m ^3-\frac{36}{Q_m ^3}+195 Q_m ^2+\frac{195}{Q_m ^2}-516 Q_m -\frac{516}{Q_m }+708\right)\nonumber\\
&+Q_\rho^2 \bigg(3 Q_m ^6+\frac{3}{Q_m ^6}-144 Q_m ^5-\frac{144}{Q_m ^5}+1572 Q_m ^4+\frac{1572}{Q_m ^4}-8304
   Q_m ^3-\frac{8304}{Q_m ^3}+25479 Q_m ^2\nonumber\\
&\hspace{1.2cm}+\frac{25479}{Q_m ^2}-48864Q_m -\frac{48864}{Q_m }+60516\bigg) \nonumber \\
&+ {\cal O}(Q_\rho^3) \,,
\end{align}
which matches with a corresponding expansion of $g^{0,(2,2,2)}$.

These very non-trivial checks lend strong support to our conjecture~(\ref{ConjectureCon}).


\section{Elliptic Genera and Topological Invariants}\label{Sect:EllGen}

In the previous sections, we studied the properties of the NS-limit of the free energy of M-strings with a compact transverse direction. We found evidence that these functions are related to the affine ${A}_{N-1}$ relative monopole string moduli space $M_{k_1,\ldots,k_{N}}$ with charges $(k_1,\cdots,k_N)$. Here, following \cite{Hohenegger:2015cba,Haghighat:2015coa}, we conjecture a concrete relation between the NS-limit of the free energy and the elliptic genus $\chi_{\text{ell}}(M_{k_1,\ldots,k_N})$ of $M_{k_1,\ldots,k_N}$ as
\bea\label{conjecture}
\chi_{\text{ell}}(M_{k_{1}\cdots k_{N}})=\left\{\begin{array}{lcl}  \frac{1}{N}\lim_{\epsilon_{2}\mapsto 0}\frac{G^{(k_{1},\ldots, k_{N})}(\tau,m,\epsilon_{1},\epsilon_2)}{G^{(1)}(\tau,m,\epsilon_{1},\epsilon_2)} & \text{for} & k_1=k_2=\ldots=k_N \\[10pt]\lim_{\epsilon_{2}\mapsto 0}\frac{G^{(k_{1},\ldots, k_{N})}(\tau,m,\epsilon_{1},\epsilon_2)}{G^{(1)}(\tau,m,\epsilon_{1},\epsilon_2)} & \text{else} &  \end{array}\right.\,.
\eea

\subsection{The Case of Charges $(k_{1},\ldots, k_{N})=(1,\ldots,1)$}

For the charge configuration $(1,1,\ldots,1)$, we see from eq.~(\ref{simpleconfiguration}) that
\begin{align}
G^{(1,\cdots,1)}(\tau,m,\epsilon_1,\epsilon_2)=W(\tau,m,\epsilon_{1},\epsilon_2)^{N-1}\Big[G^{(1)}(\tau,m,\epsilon_{1},\epsilon_2)+(N-1)F^{(1)}(\tau,m,\epsilon_{1},\epsilon_2)\Big]\,.
\end{align}
In the NS-limit, the above expression simplifies due to eq.~(\ref{NSrelation})
\begin{align}
\lim_{\epsilon_2\to 0}\frac{G^{(1,\ldots,1)}(\tau,m,\epsilon_{1},\epsilon_2)}{G^{(1)}(\tau,m,\epsilon_{1},\epsilon_2)}=N\,W(\tau,m,\epsilon_1,\epsilon_2=0)^{N-1}\,.
\end{align}
Therefore, the elliptic genus is given by
\bea
\chi_{\text{ell}}(M_{1,\cdots, 1})=W(\tau,m,\epsilon_1,\epsilon_2=0)^{N-1}\,.
\eea

\subsection{$\chi_{y}$ Genus for $M_{k_1,\cdots,k_N}$}
In the limit $\tau\mapsto i\infty$, the elliptic genus reduces to the $\chi_{y}$ genus
\begin{align}
\chi_{y}(M_{k_1,\cdots,k_N})&:=\lim_{\tau\mapsto i\infty} \chi_{\text{ell}}(M_{k_1,\cdots,k_N})\nonumber\\
&=\lim_{\tau\mapsto i\infty}\lim_{\epsilon_2\mapsto 0}\frac{G^{(k_1,\cdots,k_N)}(\tau,m,\epsilon_1,\epsilon_2)}{G^{(1)}(\tau,m,\epsilon_1,\epsilon_2)}\,.\label{DefChiy}
\end{align}
This $\tau\mapsto i\infty$ limit can easily be computed for the partition function using the results of \cite{Hohenegger:2013ala}. It is given by
\begin{align}
\lim_{\tau\mapsto i\infty}\mbox{PLog}\,\widetilde{\mathcal{Z}}^{(2)}&(\tau,m,t_{f_1},\ldots,t_{f_{N}},\epsilon_{1,2})\nonumber\\
&=\frac{N(Q_{m}+Q_{\rho}Q_{m}^{-1})-Q_{\rho}\Big(\sqrt{qt}+\frac{1}{\sqrt{qt}}+(N-1)(\sqrt{\frac{t}{q}}+\sqrt{\frac{q}{t}})\Big)}{(1-Q_{\rho})(q^{\frac{1}{2}}-q^{-\frac{1}{2}})(t^{\frac{1}{2}}-t^{-\frac{1}{2}})}\nonumber\\
&+\sum_{1\leq a<b\leq N}
\frac{(Q_{ab}+Q_{\rho}Q_{ab}^{-1})(Q_{m}+Q_{m}^{-1})-(Q_{ab}+Q_{\rho}Q_{ab}^{-1})(\sqrt{\frac{t}{q}}+\sqrt{\frac{q}{t}})}{(1-Q_{\rho})
(q^{\frac{1}{2}}-q^{-\frac{1}{2}})(t^{\frac{1}{2}}-t^{-\frac{1}{2}})}\,,
\end{align}
where we recall the definitions $Q_{ab}=Q_{f_{a}}Q_{f_{a+1}}\ldots Q_{f_{b-1}}$ and $Q_\rho=e^{2\pi i \rho}$.
Following (\ref{DefChiy}), we further need to divide by $\lim_{\tau\to i\infty}G^{(1)}(\tau,m,\epsilon_1,\epsilon_2)$ and obtain
\begin{align}
\lim_{\epsilon_2\mapsto 0}\lim_{\tau\mapsto i\infty}&\frac{\mbox{PLog}\,\widetilde{\mathcal{Z}}^{(2)}(\tau,m,t_{f_1},\ldots,t_{f_{N}},\epsilon_{1,2})}{G^{(1)}(\tau,m,\epsilon_{1,2})}\nonumber\\
&=
N\frac{Q_{m}^2}{(1-Q_{m}q^{\frac{1}{2}})(1-Q_{m}q^{-\frac{1}{2}})}+N\sum_{k\geq 1}Q_{\rho}^{k}+
\sum_{1\leq a<b\leq N}
\frac{(Q_{ab}+Q_{\rho}Q_{ab}^{-1})}{(1-Q_{\rho})}\,.
\end{align}
From this, it follows that\\
\begin{tcolorbox}
${}$\\[-30pt]
\bea
\chi_{y}(M_{k_{1}\cdots k_{N}})=\begin{cases}
1\,,\,\,\, (k_{1},\cdots, k_{N})= (k,\cdots,k)\,,\,\,k\geq 1\,\\
1\,,\,\,\, (k_{1},\cdots,k_{N})=(k,\cdots,k,k+1,\cdots k+1,k\cdots k)\,,\,\,k\geq 0\\
0\,,\,\,\,\mbox{otherwise}\,.
\end{cases}
\eea
\end{tcolorbox}
\noindent
This implies that
\begin{align}
\sum_{q=0}^{d}(-1)^{q}\mbox{dim}_{\mathbb{C}}H^{p,q}(M_{k_1\cdots k_N})=\begin{cases}
\delta_{p,0}\,,\,\,\,\,(k_{1},\cdots, k_{N})= (k,\cdots,k)\,,\,\,k\geq 1\,\\
\delta_{p,0}\,,\,\,\, (k_{1},\cdots,k_{N})=(k,\cdots,k,k+1,\cdots k+1,k\cdots k)\,,\,\,k\geq 0\\
0\,,\,\,\,\mbox{otherwise}\,,
\end{cases}
\end{align}
where $d=\mbox{dim}_{\mathbb{C}}M_{k_1,\ldots, k_N}$. The cases where some of the $k_i$ are zero, capture the $\chi_y$ genus of non-compact configurations that we studied in \cite{Hohenegger:2015cba}. For nonzero $k_i$'s, to the best knowledge of the authors, the above results for the $\chi_y$ genus are new.  It would be interesting to confirm them by a direct computation of the multi-monopole moduli space.
 
\section{Conclusions and Future Directions}\label{Sect:Conclusions}
In this paper, we have studied aspects of BPS excitations in M5-M2-brane configurations where a transverse direction is compactified. Following our previous work \cite{Hohenegger:2015cba}, these configurations allow two dual descriptions, namely in terms of M-strings and monopole-strings. A key feature of this compact setup is a manifest $SL(2,\mathbb{Z})\times SL(2,\mathbb{Z})$ symmetry (which reduces to a single $SL(2,\mathbb{Z})$ in the decompactification limit). These two modular symmetries are associated with two dual gauge theories whose partition functions we have presented explicitly. The BPS excitations in these two five-dimensional theories can physically be interpreted as instanton particles and monopole strings, respectively.

Comparing the compact partition functions to their non-compact counterparts studied in \cite{Hohenegger:2015cba} we found an interesting relationship. Indeed, the counting function of compact BPS configurations can fully be constructed as a linear superposition of the non-compact ones. The result, as summarized by Eq.(\ref{generalconfig}), points to interesting implications for the little string theories: For IIA and IIB string theories, open and closed fundamental strings are distinct states. In particular, the closed string is not treated as a composites of open strings. However, for IIa and IIb little string theories, our `wrapping' prescription Eq.(\ref{generalconfig}) implies that the little strings can be viewed as bound-states of M-strings. 
Stated differently, for the purpose of BPS counting of IIb little strings, one only needs to know BPS excitations of the (2,0) superconformal field theory, which is just the low-energy limit of the IIb little string theory.

Furthermore, by carefully studying specific expansions of the two gauge theory partition functions mentioned above, we also discovered remarkable relations between their BPS state counting. Physically, this implies new relations between specific instanton and monopole configurations, respectively, which have not been observed in the literature so far. It will be interesting, both from physics and mathematics aspects, to further explore this observation: phrased more concretely, the question is how instantons on $\mathbb{R}^4$ are related to monopoles on $\mathbb{R}^3$ and what is its physical reason. 
Another concrete question is to understand whether the relations discussed here can be generalized to instanton configurations whose contribution to the partition function depends explicitly on $t_{f_a}$.

Generalizing our previous work \cite{Hohenegger:2015cba}, we have proposed that the compact gauge theory partition function allows to extract the elliptic genus of the relative moduli space of affine ${A}_{N-1}$ monopole strings. Based on this conjecture, by computing the corresponding $\chi_y$ genus we have extracted topological data of this moduli space. The latter are not yet known in the mathematics  literature. It would be very interesting to confirm our conjectures by independent methods. 

Finally, consequences and implications of our results on the BPS excitations in Type IIa and IIb little string theories in six dimensions is a very interesting topic, which we will relegate in a forthcoming paper \cite{ours}. 

We believe that a further exploration of M5-M2-brane configurations along the lines we have advocated in this work, will shed further light on the role of tensionless strings in the elusive  six-dimensional superconformal field theories.
\section*{Acknowledgement}
We thank Ofer Aharony, Dongsu Bak, Andreas Gustavsson, Babak Haghighat, David Kutasov, Sameer Murthy and Cumrun Vafa for helpful discussions. SH and SJR acknowledge the "Liouville, Integrability and Branes (10) and (11)"  Focus Program at the Asia-Pacific Center for Theoretical Physics for excellent collaboration environment. AI acknowledges the "2015 Simons Summer Workshop on Mathematics and Physics" for hospitality during this work.  SJR was supported in part by the National Research Foundation of Korea grants 2005-0093843, 2010-220-C00003 and 2012K2A1A9055280. The work of SH is partly supported by the BQR Accueil EC 2015. A.I. was supported in part by the Higher Education Commission grant HEC-20-2518.

\appendix
\section{Compact and Non-Compact Free Energies}\label{App:PertExpand}
\subsection{Series Expansion}\label{App:SerExpansion}
We begin with the compact free energies:
{\allowdisplaybreaks
\begin{align}
&g^{0,(2)}=\frac{1}{12}\left[(2E_2(2\tau)-E_2(\tau))\,\varphi_{-2,1}+\varphi_{0,1}\right]\,,\nonumber\\
&g^{1,(2)}=\frac{1}{288}\left[4(E_2(\tau)-E_2(2\tau))\varphi_{0,1}-\left(E_2(\tau)^2-4E_2(2\tau)^2+15E_4(\tau)-12E_4(2\tau)\right)\varphi_{-2,1}\right]\,,\\
&\nonumber\\
&g^{0,(3)}=\frac{1}{1440}\left[10\varphi_{0,1}^2+10(3E_2(3\tau)-E_2(\tau))\varphi_{-2,1}\varphi_{0,1}+(37E_4(\tau)-27E_4(3\tau))\varphi_{-2,1}^2\right]\,,\nonumber\\
&g^{1,(3)}=\frac{1}{60480}\bigg[105(E_2(\tau)-E_2(3\tau))\varphi_{0,1}^2-7(5E_2(\tau)^2-45E_2(3\tau)^2+157E_4(\tau)-117E_4(3\tau))\varphi_{0,1}\varphi_{-2,1}\nonumber\\
&\hspace{1.5cm}+\big[2592 E_6(3\tau)+1496 E_6(\tau)+7E_2(\tau)(37E_4(\tau)+270E_4(3\tau))-6237E_2(3\tau)E_4(3\tau)\big]\varphi_{-2,1}^2\bigg]\,,\\
&\nonumber\\
&g^{0,(2,1)}=\frac{1}{144}\left[\varphi_{0,1}^2+2E_2(\tau)\varphi_{0,1}\varphi_{-2,1}+(3E_4(\tau)-2E_2(\tau))\varphi_{-2,1}^2\right]\,,\nonumber\\
&g^{1,(2,1)}=\frac{\varphi_{-2,1}}{432}\left[2(E_2(\tau)^2-E_4(\tau))\varphi_{0,1}-(E_2(\tau)^3+5E_2(\tau)E_4(\tau)-6E_6(\tau))\varphi_{-2,1}\right]\,,\\
&\nonumber\\
&g^{0,(2,2)}=\frac{1}{90720}\bigg[\left(546 E_2(\tau) E_4(\tau)-672 E_2(2\tau) E_4(2\tau)-601 E_6(\tau)+832 E_6(2\tau)\right) \varphi _{-2,1}^3\nonumber\\
&\hspace{0.5cm}-21 \left(10 E_2(\tau)^2-40 E_2(2\tau)^2-9
   E_4(\tau)+24 E_4(2\tau)\right) \phi _{0,1} \phi _{-2,1}^2+105 \left(E_2(\tau)+2 E_2(2\tau)\right) \phi _{0,1}^2 \phi _{-2,1}\nonumber\\
&\hspace{0.5cm}+105 \phi_{0,1}^3\bigg]\,,\nonumber\\
&g^{1,(2,2)}=\frac{2}{3628800}\bigg[ \big(2730 E_4(\tau) E_2(\tau)^2+5875 E_6(\tau) E_2(\tau)-9893 E_4(\tau)^2\nonumber\\
&+64 E_4(2\tau) \left(242 E_4(2\tau)-105 E_2(2\tau)^2\right)-40 E_2(2\tau)
   \left(3 E_6(\tau)+184 E_6(2\tau)\right)\big) \phi _{-2,1}^3\nonumber\\
   &-50 \left(28 E_2(\tau)^3+287 E_4(\tau) E_2(\tau)-224 E_2(2\tau)
   \left(E_2(2\tau)^2+E_4(2\tau)\right)-219 E_6(\tau)+352 E_6(2\tau)\right) \phi _{0,1} \phi _{-2,1}^2\nonumber\\
&+35 \left(85 E_2(\tau)^2-100 E_2(2\tau)^2-133
   E_4(\tau)+148 E_4(2\tau)\right) \phi _{0,1}^2 \phi _{-2,1}+700 \left(E_2(\tau)-E_2(2\tau)\right) \phi _{0,1}^3\bigg]\,.\label{Expandg22}
\end{align}}
Here $E_{2k}(\tau)$ are the Eisenstein series defined as
\begin{align}
E_{2k}(\tau):=1+\frac{(2\pi i)^{2k}}{(2k-1)! \zeta(2k)}\sum_{n=1}^\infty \sigma_{2k-1}(n)\, Q_\tau^n\,,\label{DefEisenstein}
\end{align}
and $\varphi_{-2,1}(\tau,z)$ and $\varphi_{0,1}(\tau,z)$ are the standard Jacobi forms of index $1$ and weight $-2$ and $0$ respectively
\begin{align}
\varphi_{0,1}(\tau,m)=4\sum_{i=2}^4\frac{\theta_i(\tau;m)^2}{\theta_i(\tau;0)}
\qquad \text{and} \qquad
\varphi_{-2,1}(\tau,m)=-\frac{\theta_1^2(\tau;m)}{\eta(\tau)^6}\,.\label{BasicJacForms}
\end{align}
where $\theta_i(\tau,z)$ are the Jacobi theta functions and $\eta(\tau)$ the Dedekind eta-function (see \cite{Eichler} for further information).

Similarly, we can write for the non-compact coefficient functions
{\allowdisplaybreaks
\begin{align}
&f^{0,(2)}=\frac{E_2(2\tau)-E_2(\tau)}{6}\,\varphi_{-2,1}\,,\nonumber\\
&f^{1,(2)}=\frac{12E_4(2\tau)-13E_4(\tau)-3E_2(\tau)^2+4E_2(2\tau)^2}{288}\,\varphi_{-2,1}-\frac{E_2(2\tau)-E_2(\tau)}{72}\,\varphi_{0,1}\,,\\
&\nonumber\\
&f^{0,(3)}=\left[\frac{20E_2(\tau)^2+7E_4(\tau)-27E_4(3\tau)}{1440}\,\varphi_{-2,1}+\frac{E_2(3\tau)-E_2(\tau)}{48}\,\varphi_{0,1}\right]\varphi_{-2,1}\,,\nonumber\\
&f^{1,(3)}=\frac{1}{60480}\bigg[105\left[E_2(\tau)-E_2(3\tau)\right]\varphi_{0,1}^2-63\left[5(E_2(\tau)^2-E_2(3\tau)^2)+13 (E_4(\tau)-E_4(3\tau))\right]\varphi_{0,1}\varphi_{-2,1}\nonumber\\
&\hspace{1.5cm}+\big[140E_2(\tau)^3-6237E_2(3\tau)E_4(3\tau)+7E_2(\tau)(137E_4(\tau)+270E_4(3\tau))+656E_6(\tau)\nonumber\\
&\hspace{1.5cm}+2592E_6(3\tau)\big]\varphi_{-2,1}^2\bigg]\,,\\
&\nonumber\\
&f^{0,(2,1)}=\frac{E_4(\tau)-E_2(\tau)^2}{96}\,\varphi_{-2,1}^2\,,\nonumber\\
&f^{1,(2,1)}=-\frac{\varphi_{-2,1}}{576}\left[\left[E_4(\tau)-E_2(\tau)^2\right]\varphi_{0,1}+\left[E_2(\tau)^3+3E_4(\tau)E_2(\tau)-4E_6(\tau)\right]\varphi_{-2,1}\right]
\end{align}}
\subsection{Relations between Compact and Non-Compact Coefficient Functions}\label{App:RelsCompNonComp}
With the expressions above (and several others which we do not display to save space) the compact and non-compact coefficients
\begin{itemize}
\item Case $K=2$
\begin{align}
&g^{n,(2)}=f^{n,(2)}+f^{n,(1,1)}\,,&&g^{n,(1,1)}= 2\,f^{n,(1,1)}\,,&&\forall n=0,1,2,3\,.
\end{align}
\item Case $K=3$
\begin{align}
g^{n,(3)}&=f^{n,(3)}+2\,f^{n,(2,1)}+f^{n,(1,1,1)}\,,\nonumber\\
g^{n,(2,1)}&=2\, f^{n,(2,1)} +f^{n,(1,1,1)}\,,
\nonumber\\
g^{n,(1,1,1)}&=3\, [f^{n,(1,1,1)} ]\,,&&\forall n=0,1,2\,.
\end{align}
\item Case $K=4$
\begin{align}
g^{n,(4)}&=f^{n,(4)}+2\,f^{n,(3,1)}+f^{n,(2,2)}+2\,f^{,(2,1,1)}+f^{n,(1,2,1)}+f^{n,(1,1,1,1)}\,,\nonumber\\
g^{n,(3,1)}&=2\, f^{n,(2,1,1)} +2\,
f^{n,(3,1)} \,,
\nonumber\\
g^{n,(2,2)}&=2\, [ f^{n,(1,1,1,1)}+\,f^{n,(1,2,1)}+\,f^{n,(2,2)}]\,,\nonumber\\
g^{n,(2,1,1)}&=f^{n,(1,1,1,1)}+2\,f^{n,(2,1,1)}+f^{n,(1,2,1)}\,,\nonumber\\
g^{n,(1,1,1,1)}&=4\,[f^{n,(1,1,1,1)}]\,,\hspace{5cm}\forall n=0,1\,.\label{RelsK4}
\end{align}
\item Case $K=5$
{\allowdisplaybreaks \begin{align}
g^{0,(5)}&=f^{0,(5)}+2\,f^{0,(4,1)}+2\,f^{0,(3,2)}+f^{0,(2,1,2)}+2\,f^{0,(2,2,1)}+f^{0,(1,3,1)}\nonumber\\
&\hspace{1cm}+2\,f^{0,(3,1,1)}+2\,f^{0,(1,2,1,1)}+2\,f^{0,(2,1,1,1)}+f^{0,(1,1,1,1,1)}\,,\nonumber\\
g^{0,(2,1,1,1)}&=f^{0,(1,1,1,1,1)} +2\,f^{0,(1,2,1,1)} + 2\,f^{0,(2,1,1,1)} \,,\nonumber\\
g^{0,(3,1,1)}&=f^{0,(1,3,1)} + 2\,f^{0,(2,1,1,1)} + 2\,f^{0,(3,1,1)} \,,\nonumber\\
g^{0,(2,2,1)}&=f^{0,(1,1,1,1,1)} + 2\,f^{0,(1,2,1,1)} + 2\,f^{0,(2,2,1)} +f^{0,(2,1,2)} \,,\nonumber\\
g^{0,(3,2)}&=f^{0,(1,1,1,1,1)} + 2\,f^{0,(1,2,1,1)} +f^{0,(1,3,1)} + 2\,f^{0,(2,1,1,1)} + 2\,f^{0,(2,2,1)} + 2\,f^{0,(3,2)} \,,\nonumber\\
g^{0,(4,1)}&=f^{0,(2,1,2)} + 2\,f^{0,(3,1,1)} + 2\,f^{0,(4,1)} \,,\nonumber\\
g^{0,(1,1,1,1,1)}&=5\,[f^{0,(1,1,1,1,1)}] \,.
\end{align}}
\item Case $K=6$
\begin{align}
g^{0,(3,3)}&=2\, [f^{0,(1,1,1,1,1,1)}+2\, f^{0,(1,2,1,1,1)}  + f^{0,(1,2,2,1)}  + f^{0,(2,1,1,2)}  + 2 \,f^{0,(2,2,1,1)}] \nonumber\\
&\hspace{1cm} +  4\, f^{0,(2,3,1)}  + 2\, f^{0,(3,3)} \,,\nonumber\\
g^{0,(2,1,2,1)}&=2\,[2\,f^{0,(2,1,2,1)}+ f^{0,(1,1,2,1,1)}]\,.
\end{align}
\end{itemize}
As we can see, to each order, we can express the compact free energies as particular linear combinations of the non-compact ones. However, these relations are not invertible, due to the fact that the compact $g^{n,(\{k_i\})}$ are invariant under cyclic rotations of the $k_i$, while the non-compact ones $f^{n,(\{k_i\})}$ are only invariant under mirror reflection.


\end{document}